%% file: sp.tex
\documentclass[conference,compsoc]{IEEEtran}
\ifCLASSOPTIONcompsoc
  \usepackage[nocompress]{cite}
\else
  \usepackage{cite}
\fi
\ifCLASSINFOpdf
\else
\fi
\usepackage{tikz}
\usepackage{amsmath}
\usepackage{filecontents}
\usepackage{multirow}
\usepackage{hyperref}
\usepackage{cleveref}
\usepackage[normalem]{ulem}
\usepackage{mathtools}
\usepackage{textcomp,amssymb}
\usepackage{setspace}
\usepackage{latexsym,fancyhdr,url}
\usepackage{enumerate}
\usepackage{amsmath}
\usepackage[ruled,vlined,linesnumbered]{algorithm2e}
\usepackage{algorithm2e}
\usepackage{algpseudocode}
\usepackage{booktabs}
\usepackage{makecell}
\usepackage{graphics}
\usepackage{xparse}
\usepackage{xspace}
\usepackage{marvosym}
\usepackage{multirow}
\usepackage{csvsimple}
\usepackage{balance}
\usepackage{makecell}
\usepackage{pifont}
\usepackage{xcolor}
\usepackage{colortbl}
\usepackage{placeins}
\usepackage[most]{tcolorbox}
\usepackage{wasysym}
\usepackage{longtable}
\usepackage{listings}
\usepackage{caption}
\usepackage{subcaption}
\definecolor{darkblue}{rgb}{0.0, 0.0, 0.55}
\definecolor{darkcandyapplered}{rgb}{0.64, 0.0, 0.0}
\usepackage{breakurl}
\usepackage{hyperref}

\DeclareCaptionType{listing}[Listing][List of Listings]

\hypersetup{
    colorlinks=true,
    linkcolor=darkblue,
    citecolor=darkcandyapplered,
    filecolor=magenta,      
    urlcolor=darkblue,
}

\useunder{\uline}{\ul}{}

\definecolor{diffadded}{RGB}{3,133,0}
\definecolor{diffremoved}{RGB}{162,0,0}
\definecolor{diffhunk}{RGB}{140,4,125}

\lstdefinelanguage{diff}{
    morecomment=[f][\bf\ttfamily\color{diffhunk}]{@@},
    morecomment=[f][\color{diffremoved}]{-},
    morecomment=[f][\color{diffadded}]{+},
    basicstyle=\ttfamily\footnotesize,
    keepspaces=true,
    showstringspaces=false,
    numbers=left,
    numberstyle=\tiny,
    numbersep=5pt,
    xleftmargin=11pt,
    breaklines=true
}

\definecolor{keywordcolor}{HTML}{b00040}
\definecolor{keywordcolor2}{HTML}{008000}
\definecolor{typecolor}{rgb}{0.17,0.57,0.68}
\definecolor{functioncolor}{HTML}{1818ff}
\definecolor{green}{rgb}{0,0.6,0}
\definecolor{red}{rgb}{0.6,0,0}
\definecolor{gray}{rgb}{0.5,0.5,0.5}
\definecolor{commentcolor}{RGB}{62,122,122} 

\lstdefinestyle{mystyle}{
    language=C,
    backgroundcolor=\color{white},   
    commentstyle=\color{commentcolor},
    keywordstyle=\color{keywordcolor},
    numberstyle=\tiny,
    stringstyle=\color{red},
    basicstyle=\footnotesize\ttfamily,
    breakatwhitespace=false,         
    breaklines=true,                 
    captionpos=b,                    
    keepspaces=true,                 
    numbers=left,                    
    numbersep=5pt,             
    xleftmargin=11pt,     
    showspaces=false,                
    showstringspaces=false,
    showtabs=false,                  
    tabsize=2,
    deletekeywords={void,char,int,if,else,for,return,struct,sizeof},
    classoffset=0,
    morekeywords={void,char,int,uint64_t},
    keywordstyle=\color{keywordcolor}\bfseries,
    classoffset=1,
    morekeywords={if,else,for,return,struct,sizeof},
    keywordstyle=\color{keywordcolor2}\bfseries,
    classoffset=2,
    morekeywords={int,char},
    keywordstyle=\color{typecolor}\bfseries,
    classoffset=3,
    morekeywords={process_input,handle_input},
    keywordstyle=\color{functioncolor},
    classoffset=4,
    morekeywords={NULL},
    keywordstyle=\color{keywordcolor2},
    classoffset=0,
}

\newcommand{\F}{\textsc{Fixmorph}\xspace}
\newcommand{\T}{\textsc{TSBPORT}\xspace}
\newcommand{\sys}{\textsc{PortGPT}\xspace}

\newcommand{\RomanNumeralCaps}[1]{\MakeUppercase{\romannumeral #1}}

\makeatletter
\renewcommand\paragraph{\@startsection{paragraph}{4}{\z@}%
  {0.1\baselineskip }%
  {-0.5em}
  {\normalfont\normalsize\bfseries}}
\makeatother

\newcommand{\lstref}[1]{\hyperref[#1]{Listing~\ref*{#1}}}

\newcommand{\listingref}[1]{\hyperref[#1]{Listing~\ref*{#1}}}

\def\Snospace~{\S{}}

\newcommand{\sref}[2]{\hyperref[#2]{#1 \ref{#2}}}
\newcommand{\ie}{\mbox{\it{i.e.,\ }}}
\newcommand{\eg}{\mbox{\it{e.g.,\ }}}

\begin{document}
%
\date{}

\title{\Large \bf \sys: Towards Automated Backporting Using Large Language Models}

\author{
Zhaoyang Li\IEEEauthorrefmark{2}\IEEEauthorrefmark{1}, 
Zheng Yu\IEEEauthorrefmark{3}\IEEEauthorrefmark{1}, 
Jingyi Song\IEEEauthorrefmark{2}, 
Meng Xu\IEEEauthorrefmark{5}, 
Yuxuan Luo\IEEEauthorrefmark{6}, 
Dongliang Mu\IEEEauthorrefmark{2}\IEEEauthorrefmark{4}\textsuperscript{\Letter}\\
\IEEEauthorrefmark{2}School of Cyber Science and Engineering, Huazhong University of Science and Technology, China\\
\IEEEauthorrefmark{2}Hubei Key Laboratory of Distributed System Security\\
\IEEEauthorrefmark{3}Northwestern University, 
\IEEEauthorrefmark{5}University of Waterloo,
\IEEEauthorrefmark{6}Canonical Ltd.,
\IEEEauthorrefmark{4}JinYinHu Laboratory, China\\
\textit{\{lizy04, jingyisong, dzm91\}@hust.edu.cn}\\
 \textit{
zheng.yu@northwestern.edu,  
meng.xu.cs@uwaterloo.ca,
yuxuan.luo@canonical.com}
}

\maketitle
\makeatletter
\long\def\@makefntext#1{\parindent 1em\noindent #1}
\makeatother
\footnotetext{\IEEEauthorrefmark{1} The first two authors contributed equally (alphabetical order).}
\footnotetext{\textsuperscript{\Letter} Corresponding author}

\input{abstract}
\input{sections/1_intro.tex}
\input{sections/2_background.tex}
\input{sections/3_motivation}
\input{sections/4_design.tex}
\input{sections/5_impl.tex}
\input{sections/6_eval.tex}

\input{sections/7_discussion.tex}
\input{sections/8_related.tex}
\input{sections/9_conclusion.tex}

\ifCLASSOPTIONcompsoc
  \section*{Acknowledgments}
  This work was supported by the National Natural Science Foundation of China (No. 62102154), State Key Lab of Processors, Institute of Computing Technology, CAS under Grant No. CLQ202301.
\else
  \section*{Acknowledgment}
\fi

{
    \footnotesize
    \bibliographystyle{plain}
    \bibliography{ref}
}

\appendices
\input{sections/10_appendix.tex}

\input{sections/11_meta_review}




%



\end{document}

%% file: abstract.tex
\begin{abstract}

Patch backporting,
the process of migrating mainline security patches to older branches,
is an essential task in maintaining popular open-source projects
(e.g., Linux kernel).
However, manual backporting can be labor-intensive,
while existing automated methods,
which heavily rely on predefined syntax or semantic rules,
often lack agility for complex patches.

In this paper,
we introduce \sys,
an LLM-agent for end-to-end automation of patch backporting
in real-world scenarios.
\sys enhances an LLM with tools to
access code on-demand,
summarize Git history, and
revise patches autonomously based on feedback (e.g., from compilers),
hence, simulating human-like reasoning and verification.
\sys achieved an 89.15\% success rate on existing datasets (1815 cases),
and 62.33\% on our own dataset of 146 complex cases,
both outperforms state-of-the-art of backporting tools.
%
%
We contributed 9 backported patches from \sys to the Linux kernel community
and all patches are now merged.

\end{abstract}

%% file: sections/1_intro.tex
\section{Introduction}



Large-scale open source projects
(\eg Linux kernel, Node.js, Debian, PostgreSQL, Kubernetes)
tend to maintain mainline,
stable, and Long-Term Support (LTS) branches
to ensure stability, continuous feature delivery, and
long-term maintenance~\cite{linuxkernel,nodejs}.
When bugs are discovered in a project,
developers tend to fix them in the mainline branch first,
after which patch backporting is performed to
retrofit a patch to stable and LTS branches.
However,
patch backporting is complex and labor-intensive~\cite{tan2022understanding,chakroborti2022backports}.
It requires maintainers to
manually resolve conflicts and
adapt patches to out-of-sync branches,
even downstream projects,
which can be both time-consuming and error-prone.



Recent years have witnessed several proposals on
automated patch backporting,
which typically consists of two stages:
\textbf{localization}
(figuring out the right location to apply code changes) and
\textbf{transformation}
(adapting code changes to be compatible with the older version).

Localization techniques have evolved throughout the years from
exact surrounding-text matching
(\eg as shown in the \href{https://www.gnu.org/software/diffutils/manual/html_node/Inexact.html}{\texttt{patch}} utility), to
predicates over nodes in typed abstracted syntax tree (AST)
(\eg \F \cite{fixmorph}), and further to
similarities in semantics-bearing program dependency graphs (PDG)
(\eg \T \cite{tsbport}).
While each advancement allows more ``fuzzy'' contexts
to be matched for hosting backported patches,
existing solutions still fall short
when code in the mainline and target branch
undergoes changes beyond what the rules/heuristics are designed for.
%
And such changes can be as simple as
renaming a function or
adjusting the location of a code snippet
(shown in~\autoref{sec:motivation})
or even declaring a buffer on stack instead of heap
(\eg \autoref{ss:bg-case-llm})



Similar to localization,
patch transformation has also evolved significantly from
text drop-in (\eg \texttt{patch}), to typed-AST based
transformation rules (\eg \F) and
semantics-oriented predefined transformation operators (\eg \T).
However,
existing works still lack the flexibility
to adapt or improvise code modification for older branches,
especially when the patch transformation does not fall into the predefined templates (in the case of \T) or rule synthesis search space (in the case of \F),
such as accurately aligning symbol names
(\eg function call, struct member, header file)
between two branches.
In addition, existing approaches are tightly coupled
with the syntactic and semantic structures of specific programming languages,
which makes it impractical to directly apply these tools
to backport patches in other languages.

Nevertheless,
while evolving localization and transformation techniques
draw inspiration from experience of human experts,
when human developers backport a patch,
they rarely follow an algorithmic procedure to locate patching points
nor do they run an exhaustive search on known patterns to adapt the patch.
Instead, human developers could perform
a mix of activities that may resemble on the lines of:
1) tracing patching location through version control metadata or symbol locator,
2) comprehending patch context changes, and
3) improvising modification to the patch based on the comprehension,
and last but not least,
4) trial-and-error.
Naturally, one direction of improving backporting tools is to
make the tool behave more like human experts.


The recent rise of large-language models (LLMs)
shed lights on how we can equip a backporting tool with
the capability of code comprehension and code generation.
In fact,
Mystique\cite{wu2025mystique}
and 
PPathF~\cite{pan2024automating} has already shown that
an off-the-shelf LLM can transform a patch
(denoted as function before and after the patch $f_o \rightarrow f_n$)
to a new context ($f_o'$)
via in-context learning~\cite{zhao2023survey, brown2020language} only
given common instructions ($I$):
\[
\text{Query}(I, f_o \rightarrow f_n, f'_o \rightarrow \,?),
\text{where}\, ? \,\text{is the LLM response}
\]
However,
they are not end-to-end backporting systems yet
as it still requires developers to manually designate
the hosting function ($f_o'$) for the ported patch.
More importantly,
they completely forgo rich information
(e.g., Git commit history, test cases, etc)
that can and will be leveraged in manual backporting.
Evidently,
they are unable to
reflect on or revise incorrect patches.
%
%


In this work,
we propose \sys,
an end-to-end LLM-based patch backporting tool.
We intentionally design \sys
by shadowing how human developers perform backporting.
%
By atomizing, analyzing, and summarizing
key actions taken by developers
as well as the information derived from the actions,
we note that developers employ a variety of capabilities,
such as
aggregating Git diffs and
counterexample-guided refinement
(see~\autoref{sec:motivation})
in backporting.
This implies an agentic architecture---%
which empowers an LLM more tools and freedom of reasoning---%
might be more suitable for backporting tasks
than in-context learning.

%
%
%

Therefore,
the key design philosophy of \sys is to
provide tools and chain-of-thoughts that we believe to be useful
(based on manual backporting experience)
to an LLM agent to facilitate its reasoning.
In terms of \textbf{localization},
we provide tools for an LLM agent to locate symbols,
selectively access relevant source code, and
Git history.
These tools help the LLM agent focus on the pertinent code and
better track code changes across branches.
%
For \textbf{transformation},
\sys allows LLM to reason and improvise based on
its knowledge base and the concrete backporting task,
but also provide best-effort validation of the generated patches and
feedback on why a patch cannot be compiled or applied
(\eg missing header file or test case failure).
We built \sys upon GPT-4o\cite{gpt-4o} and
evaluated \sys on large-scale datasets from prior works, including
1465 Linux kernel CVEs from \T and
350 Linux kernel bugs from \F.
\sys outperformed both \F and \T
with an overall success rate of 89.15\%.
%
Additionally,
we created a more complex and diverse dataset
consisting of 146 cases
sourced from 34 programs across three popular languages (C, C++, and Go).
%
On this more challenging dataset,
\sys successfully backported 62.33\% of the patches,
significantly outperforming \F and \T
(by 56.53\% and 26.09\%, respectively).

To evaluate \sys's  applicability in real-world,
we selected patches for Linux LTS and Ubuntu introduced
after the knowledge cutoff (October 2023) of GPT-4o.
On the Linux 6.1-stable branch,
\sys successfully backported 9 patches out of 18,
all of which were thoroughly verified and
subsequently accepted by the Linux community.
For Ubuntu, we tested 16 patch pairs corresponding to 10 CVEs across multiple versions, and \sys successfully backported 10 of them.

\paragraph{Summary}
This work makes the following contribution:

\textbullet~We propose a practical LLM-based backporting framework that shadows manual patch backporting workflow and leverages commonly used developer tools, such as \texttt{git}, to automatically backport patches to target versions.

\textbullet~We conduct a thorough evaluation of \sys using 1,961 patches,
    demonstrating that \sys achieves superior performance
    compared to previous works while maintaining comparable efficiency.
    
\textbullet~We leverage \sys to handle real-world patches that are difficult to backport. Nine of these patches are accepted by the Linux kernel community, demonstrating its practical applicability.


\noindent

We have open-source \sys at \url{https://github.com/OS3Lab/patch-backporting} and our dataset at \url{https://github.com/OS3Lab/patch_dataset}.


%% file: sections/2_background.tex
\section{Why LLMs for Backporting?}
\label{s:bg-why-llm}

Although LLMs have demonstrated exceptional code processing capabilities
in a diverse set of scenarios~\cite{xia2024fuzz4all, yang2025kernelgpt, wadhwa2024core, mundler2024swt},
the fundamental reason
why LLMs should be considered in backporting is beyond that.
In this paper, we argue that
the inherent context matching and code transformation schemes in LLMs,
while a blackbox in terms of explainability,
have the potential to outperform even the most comprehensive set of
syntactic or semantic rules or heuristics
defined in prior backporting works~\cite{fixmorph,tsbport},
especially with the right prompting tactics
inspired by how human experts solve backporting tasks.

This section presents a series of crafted backporting cases to illustrate
how solutions to the two key challenges in
backporting---locating matching context and transforming code patch---have evolved
from text-based to syntax-based, semantics-based, and finally
why LLMs as generative models have a unique advantage over prior rule-based designs.

\subsection{Problem Description}

Backporting,
denoted as $B: P_n \to P_o$,
is a program repair technique that adapts a patch $\Delta P$,
originally designed for a newer software version $P_n$,
to an older version $P_o$.
This approach is essential for
preserving the security and functionality of legacy systems
that cannot be fully updated due to constraints such as
compatibility issues or custom configurations.
Unlike conventional program repair practices that
rely on proof-of-concept (PoC) exploits~\cite{extractfix} or
static analysis reports~\cite{SAVER}
to produce the patch $\Delta P$,
backporting has access to $\Delta P$ already
from the beginning.
Instead,
the core challenges of backporting lies in two aspects:

1) Given $\Delta P$ is localized on code context $L_n$ in $P_n$,
we need to accurately locate $L_o$ in $P_o$ that matches $L_n$
and is suitable to host the backported patch.
        
2) We need to identify a transformation $T(\Delta P, L_o, L_n)$
such that the resulting backported patch
integrates seamlessly into $L_o$,
accounting for the differences between $L_n$ and $L_o$,
eliminating the vulnerability addressed by $\Delta P$, and
preserving functionalities of $P_o$.

        

\noindent
\textbf{Stage Setting:}
suppose we have a simple buffer-overflow vulnerability
around code context $L_n$ (\lstref{lst:bg-bug-pn})
and the bug is recently patched with $\Delta P$
shown in~\lstref{lst:bg-patch-pn}.
In the rest of this section,
we present how backporting tools have evolved
in terms of locating $L_o$ and transforming $\Delta P$.

\subsection{Vanilla Text-based Backporting}

Backporting is almost trivial when there are
no (or minimal) changes in the code context
where $\Delta P$ is applied onto (i.e., $L_n$).
For example,
in an older version $P_o$
where the \texttt{process\_input} function resides
in the same file spanning over (almost) the same lines with
identical function body shown in~\lstref{lst:bg-bug-pn}.
In this case, $L_o \approx L_n$ and
the transformation of $\Delta P$ can be handled by
the GNU \texttt{patch} utility or
\texttt{git} \texttt{cherry-pick} gracefully.

\subsection{Syntax-based Backporting}

The vanilla text-based backporting will stop working
both in locating $L_o$ and transforming $\Delta P$
when there are even simple syntatic changes,
as shown in~\lstref{lst:bg-bug-po-1}.

\begin{figure}[t]
\vspace{-1ex}
\begin{lstlisting}[style=mystyle]
void process_input(char *input) {
  char buffer[64];
  strcpy(buffer, input);
  printf("Truncated input: %s\n", buffer);
}
\end{lstlisting}
\vspace{-1.5ex}
\captionof{listing}{Code snippet in $P_n$ with a buffer-overflow}
\label{lst:bg-bug-pn}
\vspace{-2ex}
\end{figure}

\begin{figure}[t]
\begin{lstlisting}[language=diff]
@@ -1,5 +1,6 @@
 void process_input(char *input) {
   char buffer[64];
-  strcpy(buffer, input);
+  strncpy(buffer, input, 63);
+  buffer[63] = '\0';
   printf("Truncated input: %s\n", buffer);
 }
\end{lstlisting}
\vspace{-1.5ex}
\captionof{listing}{Patch $\Delta P$
to fix the bug in $P_n$ (see~\lstref{lst:bg-bug-pn})}
\label{lst:bg-patch-pn}
\vspace{-2ex}
\end{figure}

In this backporting task,
the function was named \texttt{handle\_input}
and the buffer was named \texttt{buf} in $P_o$
and somewhere along the development from $P_o$ to $P_n$,
both symbols are renamed (but the vulnerability remains).
GNU \texttt{patch} errors as it cannot find
a matching context for $\Delta P$.

\begin{figure}[h]
\vspace{-2ex}
\begin{lstlisting}[style=mystyle]
void handle_input(char *input) {
  char buf[64];
  strcpy(buf, input);
  printf("Truncated input: %s\n", buf);
}
\end{lstlisting}
\vspace{-2ex}
\captionof{listing}{$P_o$ with the same
bug shown in~\lstref{lst:bg-bug-pn}.}
\label{lst:bg-bug-po-1}
\end{figure}

Syntax-based backporting,
piloted by \F~\cite{fixmorph},
attempts to solve cases like this
by identifying matching contexts and transforming code patches
at the level of ASTs.
More specifically,
locating matching context $L_o$ is essentially checking
whether there exists an AST snippet in $P_o$ that
matches with $L_n$, generalized by heuristic rules such as
symbol names, type signatures, or even control-flow structures.
In the backporting task of~\lstref{lst:bg-bug-po-1},
\F is able to conclude that \texttt{handle\_input} is
the matching context ($L_o$) for $\Delta P$ in~\lstref{lst:bg-patch-pn}
due to AST-based fuzzy match
(see details in ~\autoref{fig:bg-ast-po-pn} of Appendix).
Subsequently, \F derives the transformation rule,
also expressed on the AST level
(shown in~\autoref{fig:bg-patch-pn-ast} of Appendix),
and apply the transformation to $L_o$,
creating a backported patch in~\lstref{lst:bg-bug-po-1-patch}.
Note that in the adapted patch,
all appearances of \texttt{buffer} in $\Delta P$
are substituted with \texttt{buf} as
on both ASTs,
var-use sites refer to the actual variable object,
not its symbol.

\begin{figure}[h]
\vspace{-3ex}
\begin{lstlisting}[language=diff]
@@ -1,5 +1,6 @@
 void handle_input(char *input) {
   char buf[64];
-  strcpy(buf, input);
+  strncpy(buf, input, 63);
+  buf[63] = '\0';
   printf("Truncated input: %s\n", buf);
 }
\end{lstlisting}
\captionof{listing}{Backported patch for
$P_o$ in~\lstref{lst:bg-bug-po-1}.}
\label{lst:bg-bug-po-1-patch}
\vspace{-2ex}
\end{figure}

\subsection{Semantics-based Backporting}

While syntax-based backporting is powerful
especially when the context matching and patch transformation
rules can be expressed with typed ASTs,
it can fail when the patch requires semantics
(either about the program or the vulnerability it fixes) to backport.
\lstref{lst:bg-bug-po-2} is an example
when backporting requires the semantic knowledge of the patched vulnerability
(buffer overflow).
While syntax-based tools (e.g., \F) can
reconcile symbol changes and produce a patch,
the patch will be the same as~\lstref{lst:bg-bug-po-1-patch} and
this is a wrong patch as now the \texttt{buf} is only 32 bytes
while both
\texttt{strncpy(buf,input,63)} and
\texttt{buf[63]='\textbackslash0'}
in the patch overflow it.

\begin{figure}[h]
\begin{lstlisting}[style=mystyle]
void handle_input(char *input) {
  char buf[32]; // NOTE: length reduced
  strcpy(buf, input);
  printf("Truncated input: %s\n", buf);
}
\end{lstlisting}
\captionof{listing}{$P_o$ with the same
bug shown in~\lstref{lst:bg-bug-pn}.}
\label{lst:bg-bug-po-2}
\vspace{-2ex}
\end{figure}

To retrofit semantic information in backported patch,
two generic approaches have been proposed in prior works:

\paragraph{Templates}
\T\cite{tsbport} categorizes patches into predefined templates such as
adding sanity checks,
modifying function call arguments,
honoring def-use relations,
and use these templates to guide patch transformation.
The process involves inferring the intention of each hunk in $\Delta P$,
and transforming this hunk to $L_o$ preserving the intention.
In the case of~\lstref{lst:bg-patch-pn},
\T infers the intention of the only hunk in $\Delta P$
is to use safer string functions and null-terminate a buffer,
and subsequently transforms the patch to~\lstref{lst:bg-bug-po-2-patch}
which applies to~\lstref{lst:bg-bug-po-2} correctly.

\paragraph{Constraint-solving}
\textsc{PatchWeave}~\cite{shariffdeen2020automated}
retrofit semantics into backporting via concolic execution.
More specifically,
it summarizes the semantic effect of $\Delta P$
(preventing a buffer overflow)
and ensure that the backported patch achieves the same effect.
However, \textsc{PatchWeave} requires a PoC input
in order to collect symbolic constraints around the patch,
which is not always readily available.
And symbolic execution inherently struggle to scale for large codebases.

\begin{figure}[h]
\begin{lstlisting}[language=diff]
@@ -1,5 +1,6 @@
 void handle_input(char *input) {
   char buf[32]; // NOTE: length changed
-  strcpy(buf, input);
+  strncpy(buf, input, 31);
+  buf[31] = '\0';
   printf("Truncated input: %s\n", buf);
 }
\end{lstlisting}
\captionof{listing}{Backported patch for
$P_o$ in~\lstref{lst:bg-bug-po-2}.}
\label{lst:bg-bug-po-2-patch}
\vspace{-2ex}
\end{figure}

\subsection{LLM-based Backporting}
\label{ss:bg-case-llm}

While effective,
semantics-based backporting,
especially those relying on heuristically defined templates,
can still fail when
the tool cannot infer the intention of the patch (or a hunk in the patch),
or cannot transform a hunk as the changes in the context ($L_o$) is not
captured by any of its templates.
\lstref{lst:bg-bug-po-3},
it is only a small tweak to~\lstref{lst:bg-bug-po-2}
by allocating the \texttt{buf} on heap instead of stack,
but \T cannot produce a correct patch as
inferring the size of \texttt{malloc}-ed object
is not a template in \T.
However,
backporting $\Delta P$ via any modern LLM
(e.g., ChatGPT 4o) is simple
even with the most obvious prompt,
as shown in the \href{https://chatgpt.com/share/6837c3d9-830c-8005-9f84-226895ce6809}{shared conversation}.
Based on this experience,
we believe LLM-based backporting can be a catch-all solution,
especially for changes beyond what is describable by heuristics and rules
in syntax- and semantics-based backporting practices.
This stance, to the best of our knowledge,
is not highlighted in prior works yet.

\begin{figure}[h]
\vspace{-2ex}
\begin{lstlisting}[style=mystyle]
void handle_input(char *input) {
  char *buf = malloc(32);
  if (buf == NULL) { return; }
  strcpy(buf, input);
  printf("Truncated input: %s\n", buf);
  free(buf);
}
\end{lstlisting}
\captionof{listing}{$P_o$ with the same
bug shown in~\lstref{lst:bg-bug-pn}.}
\label{lst:bg-bug-po-3}
\vspace{-2ex}
\end{figure}

\subsection{Backporting Types}

In this work,
we adopt the same categorization for backporting tasks
established in prior studies \cite{fixmorph,tsbport}.
Based on how significantly it differs from the original patch,
a backported patch is typically classified as:

\textbullet~\textbf{Type-\RomanNumeralCaps{1}} \textit{(No changes)}:
The backported patch is identical to the original patch
both in code and location,
meaning it can be cherry-picked trivially without any modifications.

\textbullet~\textbf{Type-\RomanNumeralCaps{2}} \textit{(Only location changes)}:
The backported patch requires no transformation,
it differs from the original patch only in where it is applied
(\eg line numbers or file names).

\textbullet~\textbf{Type-\RomanNumeralCaps{3}} \textit{(Syntatic changes)}:
The backported patch requires syntatic modifications only
(\eg function and variable names)
to ensure compatibility with the target version.

\textbullet~\textbf{Type-\RomanNumeralCaps{4}} \textit{(Logical and structural changes)}: More intrusive modifications are applied
(\eg adding or removing lines in the patch),
making the backported patch syntactically different
while preserving the same functionality.


%% file: sections/3_motivation.tex
\section{Prompting (In-context Learning) Is Not All}
\label{sec:motivation}


While \autoref{s:bg-why-llm} highlights
the inherent advantages of LLMs over traditional methods for patch backporting,
a critical question remains:
how can we best leverage and enhance their code understanding capabilities
to navigate the complexities of diverse backporting scenarios?
Indeed, a natural approach is to enable LLMs to emulate the processes employed by human developers during backporting. In this section, we present an example illustrating how human developers perform patch backporting. This demonstrates how we should equip LLM with a similar set of capabilities to emulate human developers effectively.
%
%

\subsection{Motivating Example}

\paragraph{CVE-2022-32250} \lstref{lst:CVE-2022-32250} presents the patch in the original version for a use-after-free (UAF) bug that caused CVE-2022-32250~\cite{linux-mainline-patch-CVE-2022-32250}. 
To address this issue, the core logic of the patch involves relocating the check statements from \texttt{nft\_set\_elem\_expr\_alloc} to an earlier stage in the process, specifically within \texttt{nft\_expr\_init}. However, this patch cannot be directly applied to \textit{5.4-stable},
and a call-for-volunteer is even advertised on the mailing list~\cite{CVE-2022-32250-patch-for-mainline-failed-to-apply-to-5.4-stable}.

\paragraph{Backport by human}
Human experts typically start with the \texttt{patch} utility first,
and then handle failed hunks on a hunk-by-hunk basis.
Unfortunately,
in this example,
neither of the two hunks can be applied directly due to context conflict,
\ie the code context that the patch involves
has differed between the original version and the target version.

\ding{182} To \textbf{locate} the first hunk in the target version,
the expert begins by finding the definition of function \texttt{nft\_expr\_init}.
This localization is identified by its semantic similarity to the patch; in this instance, a key \texttt{goto} statement, unique to the patch's logic, pinpoints the location.
\ding{183} Observing that the patch's surrounding context had evolved between versions, to \textbf{transform} the hunk, the expert then consulted git history to understand the reasons for these changes before attempting conflict resolution.
\ding{184} Subsequently, the expert examines the change history of the code snippet through \texttt{git} and resolves the namespace conflicts that \texttt{expr\_info} should be replaced with \texttt{info} in the target version.
\ding{185} with that,
the expert completes the first hunk of the patch.

\ding{186} To \textbf{locate} the second hunk in the target version,
the expert first attempts to find function \texttt{nft\_set\_elem\_expr\_alloc} 
by its symbol but fails as the function name does not exist in the old version.
\ding{187} Then, the expert traces the origin of this function in the original version with \texttt{git} and finds that it was introduced by commit \textit{a7fc93680408}.
\ding{188} Next, the expert uses \texttt{git show} to examine the details of this commit and concludes that \texttt{nft\_set\_elem\_expr\_alloc} was migrated from \texttt{nft\_dynset\_init} in another file.
\ding{189} Therefore, the expert views function \texttt{nft\_dynset\_init}
in the old version instead and makes sure it is the correct host for backported patch.
\ding{190} Finally, referencing the git history, the expert infers that \texttt{priv->expr} in the target version is equivalent to \texttt{expr} in the patch, and consequently deletes the corresponding statements.

\begin{figure}[t]
\vspace{-2ex}
\begin{lstlisting}[language=diff]
--- a/net/netfilter/nf_tables_api.c
+++ b/net/netfilter/nf_tables_api.c
@@ -2873,27 +2873,31 @@ *nft_expr_init(
  err = nf_tables_expr_parse(ctx, nla, ...);
  if (err < 0) goto err1;

+ err = -EOPNOTSUPP;
+ if (!(expr_info.ops...flags & NFT_EXPR_STATE))
+   goto err_expr_stateful;

  err = -ENOMEM;
  expr = kzalloc(expr_info.ops->size, ...);
@@ -5413,9 +5417,6 @@ *nft_set_elem_expr_alloc(
    return expr;
 
  err = -EOPNOTSUPP;
- if(!(expr->ops->type->flags & NFT_EXPR_STATE))
-   goto err_set_elem_expr;
-
  if (expr->ops->type->flags & NFT_EXPR_GC) {
    if (set->flags & NFT_SET_TIMEOUT)
      goto err_set_elem_expr;
\end{lstlisting}
\captionof{listing}{Original Patch for CVE-2022-32250}
\label{lst:CVE-2022-32250}
\vspace{-2ex}
\end{figure}

\ding{191} After completing the patch backporting for the target version,
the expert runs tests to verify
its correctness (as in no regression) and
effectiveness (as in defending against the PoC exploit, if available).
Additionally, the expert refines the patch by addressing potential issues identified during compilation, testsuite evaluations, and PoC testing.


\subsection{Observations}

Based on the manual patch backporting process presented above,
we draw some key observations that influence the design of \sys---%
our LLM-based backporting tool.

\emph{1)}
It is impossible to fit all information
possibly needed
by end-to-end backporting
(\eg the entire commit history)
into a single or even a pre-defined sequence of prompts.

\emph{2)}
Information relevant to the current step
can be retrieved frequently on an as-needed basis,
but it still needs to be presented in a concise way
in order not to overload the context limit of a human brain.

\emph{3)}
Decisions are frequently taken based on results of previous steps
in a trial-and-error manner,
\eg tracing the provenance of a symbol when symbol resolution fails while
viewing function body of lookup is successful.

\emph{4)}
Backported patch doesn't have to be generated in one-take.
Validating and subsequently improving the patch
based on feedback is an inseparable step in the manual process.





\noindent
Based these observations,
we believe an agentic design is well-suited for LLM-based backporting,
and identify three key capabilities
to retrofit a certain degree of
human expertise, common workflows, and best-practices
into the agent.

\textit{Basic code viewing and symbol lookup}: It is essential to equip LLM with basic code access capabilities, as what human expert does in steps \ding{182}\ding{183}\ding{186}\ding{189}. By utilizing symbol locating and code viewing, LLM can access the code context of the target version.

\textit{Access aggregated diff and track provenance}
: By aggregating historical commits of the patch code, we can precisely trace the location of the code change in the target version, as shown in steps \ding{184}\ding{187}\ding{188}. Whether it involves code relocation or identifying code segments absent in the target version, analyzing the modification history allows for precise determination. Moreover, the history of the changes clearly reveals the changes of the namespace present in the context. 

\textit{Counterexample-guided refinement}: Since the patch from the original version may include dependencies (\eg header file imports) unavailable in the old version, basic compilation tests are essential. If there are compilation errors, the previously generated patch can serve as a counterexample, using compiler error messages to guide the refinement of the patch. This step is also a practice typically performed by experts after developing the patch, as shown in step \ding{191}.

%% file: sections/4_design.tex
\section{Design Details of \sys}

%
In order to empower LLM agents with the capabilities presented in~\autoref{sec:motivation},
we design a series of tools and workflows
and further implement several optimizations to
standardize and improve the formatting of LLM outputs.
In this section, we describe these design details of \sys.
We start with an overview of \sys (\autoref{sec:overview}),
followed by detailed description of two main stages of \sys:
Per-Hunk Adaptation (\autoref{sec:per-hunk}) and
Final Patch Combination (\autoref{sec:combination}).

\begin{figure*}
    \centering
    \includegraphics[width=\linewidth]{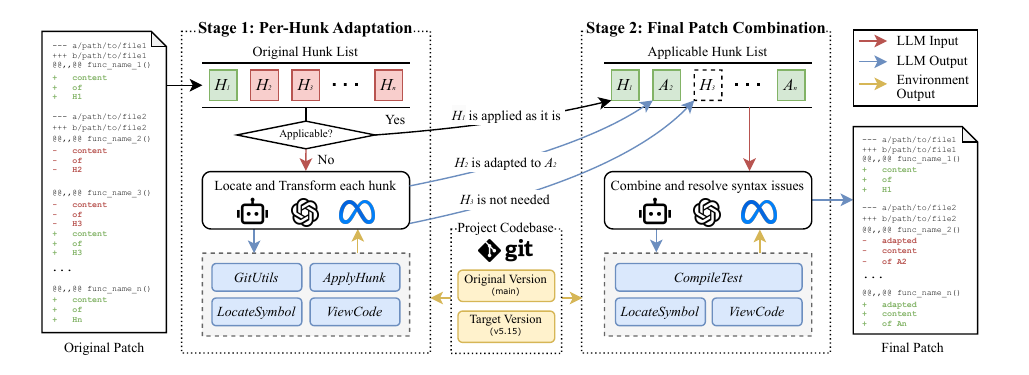}
    \caption{\textbf{Workflow of \sys}}
    \label{fig:workflow}
    \vspace{-2ex}
\end{figure*}

\subsection{Overview}\label{sec:overview}

As illustrated in \autoref{fig:workflow},
the overall workflow of \sys consists of two main stages:
Per-Hunk Adaptation and
Final Patch Combination.
In the first stage,
\sys completes \textbf{localization} and initial \textbf{transformation}
for each hunk individually.
\sys extracts hunks from the original patch and per each hunk,
\sys first determines whether the hunk requires backporting, and
if so, transforms the hunk to ensure compatibility with the target version.
This stage aims to ensure that the generated patch can be successfully applied to the target version. 
%
In the second stage,
\sys combines the transformed hunks,
applies the entire patch to the target version,
and sends the backported codebase for compilation.
If compilation failed,
\sys attempts to resolve them by
adding necessary definitions or
adjusting the code context to finalize the \textbf{transformation}.
Ultimately,
\sys outputs backported patches that are free from compilation errors.
Both stages leverage an LLM agent,
supplemented with customized tools
designed to enhance the LLM's performance.

%
%

\subsection{Per-Hunk Adaptation}
\label{sec:per-hunk}

Backporting a hunk to the target version takes two steps.
First, it is necessary to determine
if the hunk should be backported and
to identify the appropriate matching context.
Second, the hunk must be transformed
to match the host context in the target version.
To complete these steps,
\sys must gather sufficient information from the codebase to understand
how the codebase has evolved across different branches and
how the current hunk relates to the surrounding code.
%
%
To support this process,
we have designed three tools called
\textit{ViewCode}, \textit{LocateSymbol}, and \textit{GitUtils}
to facilitate information retrieval.
Once a hunk is transformed,
\sys uses another tool called \textit{ApplyHunk}
to apply the adapted hunks to the target version.
%

\textbullet~\textbf{\textit{ViewCode}.} This tool takes four parameters: \textit{ref}, \textit{file}, \textit{start}, and \textit{end}, which specify the commit, file path, and the line range to be retrieved. It returns the code snippet between the specified line numbers in the given version of the codebase, enabling the LLM to access any portion of the source code from any Git-tracked version.

\textbullet~\textbf{\textit{LocateSymbol}.} This tool takes two parameters: \textit{ref} (the specific commit) and \textit{symbol} (language objects, such as function name, variable name). It returns the definition site of the symbol in the specified version, including the file path and line number, allowing the LLM to accurately locate symbols within the codebase.

\textbullet~\textbf{\textit{GitUtils}.} This tool consists of two components: \textit{History} and \textit{Trace}, neither of them directly takes any parameters from the LLM. The \textit{History} component presents the evolution of code snippets across commits or versions, while \textit{Trace} provides detailed insights into the trace of code movement. They are designed to provide information on code evolution and commit history to the LLM. The detailed design will be discussed later.

\textbullet~\textbf{\textit{ApplyHunk}.} This tool accepts one parameter: \textit{patch} and allows LLM to apply the patch to the target version. To mitigate potential errors in LLM-generated patches, it automatically corrects patch formatting issues and generates diagnostic feedback for the LLM if the patch failed to apply. Further details are provided below.

\paragraph{\textbf{\textit{GitUtils} Detail}} The \textit{History} component displays a list of Git commits that modify the code snippet within the current hunk from the \textit{fork point} to the original version. The \textit{fork point} refers to the divergence between the original and target versions, \ie the most recent common ancestor. For example, v6.1 tag is \textit{fork point} of the mainline branch and the linux-6.1.y stable branch.
The tool analyzes the file modification history and identifies all changes made from the \textit{fork point} to the original version. It further filters the history to include only changes affecting the lines within the hunk. This design ensures that the output is both comprehensive and relevant. By focusing on commits from the \textit{fork point} onward and isolating the changes to the lines in the hunk, \textit{History} provides a detailed evolution of the code while excluding unrelated modifications.

While the component \textit{History} reveals local changes in the corresponding code snippet of a hunk, it may overlook critical
global changes, such as: the relocation of code.
%
These global changes, though important,
are challenging to incorporate directly into \textit{History}
because they are often too lengthy for an LLM.
%

To address this limitation,
we introduce \textit{Trace}
to extract and highlight significant global changes, particularly the relocation of patch contexts.
We observe that code movement changes have a direct impact on backporting. Code movement refers to the relocation of a code snippet from one location to another after the \textit{fork point}, resulting in discrepancies in its placement between the original and target versions.
To produce an accurate backported patch, LLM must identify the code movement and its target location in a patch.
Leveraging the minimal edit distance method \cite{smith1981identification}, we compare the added code with the deleted code to facilitate LLM in the code movement identification. Edit distance is an effective metric—it captures syntactic similarity and is applicable across different programming languages. Finally, with the location of the matched code and its associated commit message provided by \textit{Trace}, LLM determines if the code is migrated from that location or the code is newly introduced.

\begin{figure*}
    \centering
    \includegraphics[width=\linewidth]{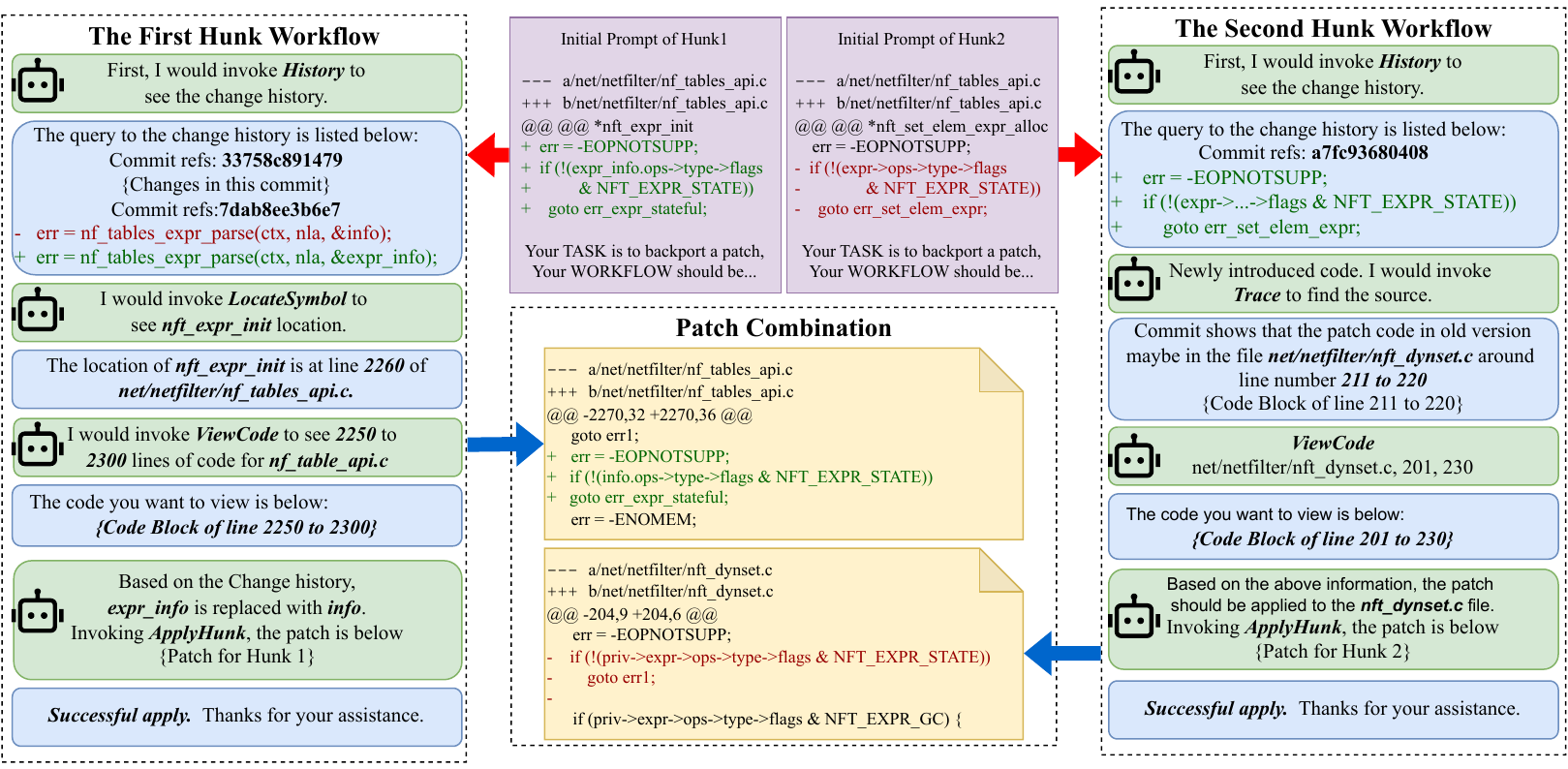}
    \caption{\textbf{Example of Per-Hunk Adaptation.}  The red arrows represent LLM agent inputs, while the blue represent LLM agent outputs. The green box represents the invocations of LLM, while the blue box represents the outputs of tools. }
    \label{fig:stage1}
    \vspace{-2ex}
\end{figure*}

\begin{algorithm}[t]
\caption{How \textit{ApplyHunk} corrects a patch and generates feedback for different failure reasons.}
\label{algorithm}
\KwIn{A patch to be corrected and applied;
A flag indicates whether to activate the Automated Context Correction mechanism.}
\KwOut{Patch apply feedback;}
\SetKwFunction{func}{apply\_feedback}
\SetKwProg{Fn}{Function}{:}{}

\Fn{\func{patch, flag}}{
    output = str()\;
    possible\_files = list()\;
    patch = format\_patch(patch); {\scriptsize\tcp{avoid "invalid patch"}}
    \If{flag}{
        patch = automated\_context\_correction(patch)\;
    }
    result = git\_apply(patch)\;

    \If{"File not found" in result}{
        old\_file = find\_rename\_file(patch)\;
        \If{old\_file}{
            \Return old\_file\;
        }
        possible\_files.append(locate\_symbol(patch))\; possible\_files.append(find\_similar\_file(patch))\;
        \For{path \textbf{in} possible\_files}{
            patch = revise\_patch(patch, path)\;
            feedback = apply\_feedback(patch)\;
            \If{"success" in feedback}{
                \Return path\;
            }
            output = output + path + feedback\;
        }
        \Return output\;
    }
    \If{"Error while searching for context" in result}{
        C\_llm = extract\_context(patch)\;
        C\_tgt = find\_corresponding\_context(C\_llm)\;
        \Return compare(C\_llm, C\_tgt)\;
    }
    \Return "success"\;
}
\end{algorithm}

\paragraph{\textit{ApplyHunk} Detail}
This tool systematically handles
errors that may arise during \texttt{git apply}.
For each type of error,
we either follow a customized correction scheme or
generate corresponding feedback to guide the LLM
for an iterative refinement of the patch (\ie the transformed hunk) as detailed in \autoref{algorithm}.

\textit{Corrupted  patch.} This error arises when the patch does not adhere to the correct format\cite{format}, \ie LLM forgets to add a space at the beginning of the code line. Therefore, the tool avoid such error by checking each code line of the patch and adding required spaces. 

\textit{Non-existent file path.} This error occurs when the patch is applied to a file that does not exist, often due to the file being renamed or the relevant function residing in a different file in the original version. To address this, the tool tries to identify possible file path lists. Initially, the tool checks if the file has been renamed. If not, the tool locates possible files in the target version by either matching the original version file name or locating one symbol referenced in the patch. Finally, the tool applies the patch to all identified possible file paths and returns apply feedback to LLM.

\textit{Context mismatch.} Patch context refers to lines that start with ` ' (space) or `-' in the patch. This error occurs when the context of the LLM-generated patch (\( C_{l} \)) is inconsistent with the target code. To resolve this, this tool identifies the most similar code block with the minimal edit distance in the target version as \( C_{t} \). Then, it compares \( C_{l} \) with \( C_{t} \) line by line and highlights any discrepancies for the LLM, allowing it to make the necessary adjustments. Furthermore, we propose an \textbf{Automated Context Correction} mechanism, which forcibly substitutes lines in \( C_{l} \) that differ from \( C_{t} \) before applying the patch.
Once the correction is made, the patch can be successfully applied. To preserve the LLM's original intent as much as possible, the tool only activates automated context correction after three failed attempts to apply the patch due to context mismatches.


\paragraph{Patch Localization}
Locating a matching context in the target version to host an adapted hunk
includes two subtasks:
identifying the correct target file path and
locating the corresponding code block within that file.
We find possible locations by applying the original hunk with \textit{ApplyHunk} and
provide possible locations in tool's output to LLM.
LLM could adopt two distinct approaches
for using \textit{GitUtils} to narrow down to the correct location.

\textit{Symbol-Based Localization:}
This workflow is particularly effective
when the patch does not involve newly introduced code or complex code movement.
In this approach,
the LLM first invokes \textit{History} to retrieve the code change history and 
determine whether the original symbol name has been renamed in the target version.
If renamed,
it then uses \textit{LocateSymbol} to identify the location of the renamed symbol within the target codebase.
Cooperated with \textit{ViewCode}, LLM further checks code blocks around those symbols and determine where the patch should be backported to.

\textit{Movement-Aware Localization:}
In more complex scenarios,
such as a code snippet has been moved to several different locations, \textit{LocateSymbol} may not be sufficient to identify the required symbols.
In such cases, the LLM invokes \textit{Trace} to determine the specific location of the code in the target version or to confirm if the code no longer exists. This ensures accurate localization and facilitates modifications even in challenging situations. 



\paragraph{CVE-2022-32250} To better understand the operations in the first stage, we continue with the patch of CVE-2022-32250 (shown in \lstref{lst:CVE-2022-32250}). The overview of this process is provided in \autoref{fig:stage1}.

Regarding the first hunk, the LLM sequentially performs the following steps: \ding{182} Invokes \textit{GitUtils}. From \textit{History}, the LLM observes, \texttt{expr\_info} has been replaced with \texttt{info} in the old version. \ding{183} Invokes \textit{LocateSymbol} to determine the location of \texttt{nft\_expr\_init} (modified by the hunk) in the old version. \ding{184} Invokes \textit{ViewCode} to inspect the code context surrounding the identified location. \ding{185} Generates the transformed patch based on the gathered information and then invokes \textit{ApplyHunk} to validate its correctness.
For the second hunk, it involves more complex scenarios. The LLM first \ding{182} invokes \textit{History}, which shows that the current code context was introduced in the \textit{a7fc93680408} commit. This satisfies the conditions for calling \textit{Trace}. Consequently, the LLM proceeds to \ding{183} invoke \textit{Trace}. Through minimal edit distance similarity matching, \textit{Trace} identifies that this code originates from Line 211 to 220 in \texttt{net/netfilter/nft\_dynset.c}. With precise code location information, the LLM then follows the same workflow and invokes \ding{184} \textit{ViewCode} and \ding{185} \textit{ApplyHunk} to complete the patch transformation. After transforming both hunks, \sys generates the combined patch.



\subsection{Final Patch Combination}\label{sec:combination}

After the inital patch is generated from previous stage, \sys will combine all the generated patches of each hunk and check for compilation errors. If there are any errors, \sys will resolve them by adding necessary definitions or adjusting the code context, which is achieved by an LLM equipped with a series of customized tools which included \textit{ViewCode} and \textit{LocateSymbol} we mentioned before with a new tool \textit{CompileTest}. Same as tool \textit{ApplyHunk}, \textit{CompileTest} takes a draft patch as input, the tool will apply the patch to the target version and compile the code. If the code can be compiled successfully, the tool will output the final patch and terminate the second stage. Otherwise, the tool will return the error message.
Typically, the compilation error messages are long and complex, which may degrade the performance of the LLM\cite{levy2024same}. To address this issue, we design a customized error message parser to extract the key information from the error message. The parser is based on a set of regular expressions that are designed to capture the most common error patterns. The parser extracts the error type, the file name, the line number, and the error message. The extracted information is then used to guide the LLM in resolving the compilation errors.

\begin{figure}[!t]
\vspace{-2ex}
\centering
\begin{subfigure}[t]{\linewidth}
\begin{lstlisting}[language=diff]
--- a/fs/cifs/cifs_debug.c
+++ b/fs/cifs/cifs_debug.c
@@ -332,6 +332,11 @@
 struct cifs_ses *ses;
 list_for_each(...) {
+   spin_lock(&ses->ses_lock);
+	if (ses->ses_status == CifsExiting) {
+		spin_unlock(&ses->ses_lock);
+		continue;
+	}
	if ((ses->serverDomain == NULL) ||
\end{lstlisting}
\caption{Patch Generated by Stage-1}
\label{subfig:CVE-2023-52752-stage-1}
\end{subfigure}

\begin{subfigure}[t]{\linewidth}
\begin{lstlisting}[language=diff]
--- a/fs/cifs/cifs_debug.c
+++ b/fs/cifs/cifs_debug.c
@@ -332,6 +332,11 @@
 struct cifs_ses *ses;
 list_for_each(...) {
+   spin_lock(&GlobalMid_Lock);
+	if (ses->status == CifsExiting) {
+		spin_unlock(&GlobalMid_Lock);
+		continue;
+	}
	if ((ses->serverDomain == NULL) ||
\end{lstlisting}
\caption{Patch Generated by Stage-2}
\label{subfig:CVE-2023-52752-stage-2}
\end{subfigure}

\caption{Patches Generated for CVE-2023-52752.}
\label{fig:CVE-2023-52752-before-stage-2}
\vspace{-2ex}
\end{figure}


To better understand how the second stage works, we provide an example. \autoref{subfig:CVE-2023-52752-stage-1} presents the patch for CVE-2023-52752~\cite{linux-patch-CVE-2023-52752} generated by Stage-1. 
%
However, the \texttt{cifs\_ses} structure (Line 4) does not have members \texttt{ses\_lock} and \texttt{ses\_status}, leading to a compilation error. \sys will issue an error message to the LLM: ``\texttt{fs/cifs/cifs\_debug.c} at Line 340, \texttt{cifs\_ses} has no member \texttt{ses\_lock}'' (\texttt{ses\_status} same), requesting a fix. Based on this information, the LLM will follow these steps:
\ding{182} First, it invokes \textit{LocateSymbol} to find the definition of \texttt{cifs\_ses}, with \textit{LocateSymbol} outputting that the symbol is defined in \texttt{fs/cifs/cifsglob.h} at Line 968. \ding{183} Then, it invokes \textit{ViewCode} to inspect the definition of \texttt{cifs\_ses}. In the definition, \texttt{ses\_status} corresponds to a similar field \texttt{status}.  The comment also indicates that \texttt{status} is protected by \texttt{GlobalMid\_Lock}. \ding{184} Based on the content from \textit{ViewCode}, the LLM deduces that \texttt{ses\_status} in the target version should be replaced by \texttt{status}. 
The semantics indicate that \texttt{ses\_status} is protected by \texttt{ses\_lock}, so \texttt{status} should be protected by \texttt{GlobalMid\_Lock} accordingly.
The LLM will then generate the revised patch in \autoref{subfig:CVE-2023-52752-stage-2} and invoke the validation tools to verify its correctness.

\subsection{Prompt Design}\label{sec:prompt}

To enable the LLM to perform the backporting task more effectively,
we carefully craft the prompt for each stage.
Agents in the two stages share the same system prompt, which explains the concept of patch backporting, the provided tools and their usage. This ensures that the LLM understands the context of patch backporting and the resources it can access. The user prompts differ between two stages since they are different tasks (shown in \autoref{fig:prompt} in Appendix). Notably, both user prompts consist of two key components: the task, which specifies what the LLM needs to accomplish, and the workflow, which provides detailed guidance on how to execute the task effectively.

\emph{In the first stage},
the objective is to generate a patch that can be successfully applied hunk by hunk. The user prompt specifies
the patch to be backported,
the original version,
and he target version and
provides a clear workflow to guide the LLM.
Specifically, LLM firstly locates where to apply the code changes in the target version by: \ding{182} utilizing similar code blocks we provided in the prompt, \ding{183} invoking \textit{GitUtils} to to analyze code change history, \ding{184} invoking \textit{LocateSymbol} to identify the locations of functions or variables.
The user prompt also instructs the LLM
to transform a hunk to align with the target version’s context,
leveraging tools such as \textit{ViewCode}
to access and analyze the target version’s code.

%
\emph{In the second stage},
the objective is to ensure that the complete patch is reliable and
does not introduce new risks.
The LLM is prompted to iteratively refine the patch based on validation feedback,
which may include compilation error messages produced by the \textit{CompileTest} tool.
In summary, the prompts are structured to empower an LLM for backporting tasks
by ensuring it knows how to use the tools to
gather as much necessary information as possible
while also following a clear and systematic workflow.

%% file: sections/5_impl.tex
\section{Implementation}
\label{sec:impl}

\sys uses LangChain \cite{langchain} to
facilitate the integration of LLM agents and prompt engineering,
and existing tools for symbol resolution and patch validation.

\paragraph{Locate Symbol} For symbol resolution, we use \textbf{ctags} \cite{universal-ctags} as the syntax parser. As a tool widely used in text editors for fast navigation, ctags supports syntax parsing for dozens of programming languages, such as C, C++, Go. In \sys, by parsing the output of ctags, we could quickly generate an available symbol table for \textit{LocateSymbol}. Specifically, ctags parses symbols such as functions, structs, and variables that appear in the codebase. It then outputs the symbol's location information in the format of ``\texttt{symbol\_name file lineno symbol\_type}''. By reading the output in the ctags format, we can generate a symbol table with symbols and their corresponding locations.

\paragraph{Context Matching} We treat lines prefixed with `-` or space in the generated patch as patch context. These are extracted and compared against all equally sized code blocks in the target file using the edit distance. The block with the minimal distance is selected, leveraging the fact that original and target code usually retain strong textual similarity despite version differences.

\paragraph{Validation Chain} To ensure that the LLM-generated patch does not introduce new security risks when both the PoC and functionality test suites are available, we design a rigorous, multi-stage validation chain following the second stage. First, we execute the original PoC on the patched version to confirm that the vulnerability has been successfully eliminated. 
And then, we run comprehensive functionality test suites to verify that the patch preserves the intended behavior and does not compromise the program's correctness. Finally, we would additionally validate a backported patch if a directed fuzzer suitable for the underlying project can be used. This layered validation strategy significantly enhances the reliability and safety of the generated patches.


%% file: sections/6_eval.tex
\begin{table*}[t]\small
    \centering
    \tabcolsep=4.5pt
    \begin{tabular}{c|cccccc}
    \toprule
    \textbf{Dataset}      
    & \textbf{System} & \textbf{Type-\RomanNumeralCaps{1}} & \textbf{Type-\RomanNumeralCaps{2}}  & \textbf{Type-\RomanNumeralCaps{3}}  & \textbf{Type-\RomanNumeralCaps{4}}  & \textbf{Total} \\ \midrule 
    \multirow{4}{*}{\begin{tabular}[c]{@{}c@{}}Prior works\\ \cite{fixmorph,tsbport}\end{tabular}} 
    
    & \F   &   20/170 (11.67\%) & 374/1208 (30.96\%) & 23/92 (25.00\%) & 30/345 (8.70\%) &  447/1815 (24.63\%)\\
    & ChatGPT$^\dagger$   &   67/170 (39.41\%) & 451/1208 (37.30\%) & 16/92 (17.58\%) & 22/345 (6.38\%) &  556/1815 (30.63\%)\\
    & \T   &  170/170 (100.00\%) & 1190/1208 (98.51\%) & 69/92 (75.00\%) & 160/345 (46.38\%) & 1589/1815 (87.59\%)      \\
    & $\T^*$   &  162/170 (95.29\%) & 919/1208 (76.08\%) & 61/92 (66.30\%) & 150/345 (43.48\%) & 1292/1815 (71.18\%)      \\
    & \sys &  170/170 (100.00\%) &   1186/1208 (98.18\%)      &   74/92 (80.43\%)     &    188/345 (54.49\%)    &    1618/1815 (89.15\%)   \\
    \midrule
    \multirow{4}{*}{Ours (C)}                                     
    &  \F      &   N/A  &   0/6 (0.00\%)   &  3/29 (10.34\%)     &  0/34 (0.00\%)    &  3/69 (4.34\%)    \\
    &  \T      &   N/A  &   5/6 (83.33\%)    &  14/29 (48.28\%)  &   5/34 (14.71\%)    &  24/69 (34.78\%)    \\
    & $\T^*$ &   N/A  &   2/6 (33.33\%)    &  14/29 (48.28\%)     &   4/34 (11.76\%)    &  20/69 (28.99\%)\\
    &  \sys    &   N/A  &   6/6 (100\%)    &  21/29 (72.41\%)     &  15/34 (44.12\%)     &  42/69 (60.87\%)     \\
    \midrule
    Ours (C++)
    &  \sys    &   N/A  &   N/A    &  10/11 (90.91\%)     &  5/17 (29.41\%)     &  15/28 (53.57\%)     \\
    \midrule
    Ours (Go)
    &  \sys    &   N/A  &   13/13 (100\%)    &  7/9 (77.78\%)     &  14/27 (51.85\%)     &  34/49 (69.39\%)     \\
    \bottomrule
\end{tabular}
    \caption{\textbf{Performance Comparison of Backporting.} The first two sections of the table compare \sys with \F and \T using datasets from prior works as well as C cases on our own dataset. The last two rows display the performance of \sys on C++ and Go cases from our dataset. ChatGPT$^\dagger$ is from \T~\cite{tsbport}. 
    \textbf{$\T^*$} represents the performance of \T under the same settings as \sys and \F. Unlike \sys and \F, \T requires the target file for each hunk as input, which may not be available in real-world scenarios. To ensure a fair comparison, we introduce \textbf{$\T^*$} as a variant of \T that operates without access to the target file, simulating a more realistic environment.}
    
    \label{table:perf}
    \vspace{-2ex}
\end{table*}

\section{Evaluation}


In this section, we assess \sys to address the following research questions:

\textbullet~\textbf{RQ1 (Performance)}: How does \sys compare to other works in terms of the success rate of backporting?

\textbullet~\textbf{RQ2 (Ablation)}: How does each module in \sys contribute to its overall performance?

\textbullet~\textbf{RQ3 (Efficiency)}: How well \sys handles the backporting task in terms of time and cost efficiency?

\textbullet~\textbf{RQ4 (Practicality)}: How applicable is \sys for handling backporting tasks in real-world scenarios?

%


\subsection{Settings}
\paragraph{Environment} All experiments were conducted on a 64-bit Ubuntu 24.04 LTS server equipped with an Intel(R) Xeon(R) Gold 6248R 96-core CPU running at 3.00 GHz, 512 GB of memory, and a 22 TB hard drive. The large language models used in our evaluation include
GPT-5 \cite{gpt-5} (version \textit{gpt-5}),
GPT-4o \cite{gpt-4o} (version \textit{gpt-4o-2024-08-06}), Gemini-2.5-Flash \cite{gemini-2.5-flash} (version \textit{gemini-2.5-flash}),
Llama 3 \cite{llama3} (version \textit{Llama 3.3 70B}) and 
DeepSeek-v3 \cite{deepseekai2024deepseekv3technicalreport} (version \textit{DeepSeek-V3-0324}).

\paragraph{Datasets} We evaluate \sys using 1961 patches, which include 350 patches from \F\cite{fixmorph}, 1465 patches from \T\cite{tsbport}. And 146 patches collected by ourselves (details in \autoref{tab:dataset} in Appendix).
These patches span 34 programs across three languages (C, C++, and Go), with the number of modified lines ranging from 1 to 1792.
%
These 146 patches are collected based on the following criteria:
\ding{182} The corresponding vulnerability of the patch
is recorded in the CVE database within the past four years.
\ding{183} At least one version within the same project
has not reached end-of-life and requires backporting.
\ding{184} The patch cannot be trivially applied to the target version,
\ie not Type-\RomanNumeralCaps{1}.

\paragraph{Evaluation Criteria}  


We compare \sys with \F and \T using their datasets
(1,815 cases in total)
along with all C cases in our dataset.
Comparisons with C++ and Go cases from our dataset were excluded,
as both tools support C only.
Although we attempted to adapt these tools for use with C++ and Go,
their implementations are highly tailored to C,
making migration infeasible. 

Additionally,
unlike \sys and \F,
\T requires the target file for each hunk as input---%
a condition that assumes some degree of ground-truth patch localization,
which is not entirely realistic.
To ensure a fair comparison, we introduce a variant of \T, denoted as $\T^*$, that operates under the same settings as \sys and \F. $\T^*$ is designed to function without access to the target file, thereby simulating a more realistic environment. By aligning the evaluation conditions, we enable a more practical and meaningful comparison of the performance among the three tools.

We excluded \textsc{SkyPort}~\cite{shi2022backporting} from our comparison,
as it is designed for web applications (PHP only),
which is not included in our dataset.
Mystique\cite{wu2025mystique}
and \textsc{PPatHF}\cite{pan2024automating} was excluded
as it is designed for function-level patch porting and
requires perfect localization,
whereas backporting typically involves multiple functions.
Although we attempted to compare \textsc{PatchWeave}~\cite{shariffdeen2020automated},
its dependence on KLEE symbolic execution presents challenges.
The use of symbolic execution can result in path explosion,
which makes it impractical for large-scale programs (\eg Linux kernel).
Furthermore,
\textsc{PatchWeave} requires PoC for the corresponding vulnerabilities,
which is not available in our dataset.





\paragraph{Validation Methodology}
To validate the correctness of the patches generated by \sys,
we compare them with the ground truth.
If the patches are not identical,
we conduct human validation to ensure their correctness,
following best practices outlined in \cite{fixmorph,tsbport}.
%
%
Each patch is independently reviewed by three researchers
to guarantee accuracy and consistency.
We also try our best to collect PoC and test suites
for each vulnerability in our dataset,
following the validation chain described in~\autoref{sec:impl}.
%
%
To evaluate whether \sys-generated patches
introduce new security vulnerabilities,
we run directed fuzzing (SyzDirect~\cite{Xin2023SyzDirect}) on
successfully backported Linux patches that differ from ground-truth.
%
%
No security risks were identified during 4 hours of fuzzing.

\subsection{Performance Evaluation (RQ1)}\label{sec:performance}


\autoref{table:perf} provides a summary of \sys's performance, which is based on GPT-4o, in comparison to \F and \T, while
%
%
\autoref{table:models} illustrates \sys's performance
when powered by different LLMs. 

%


\subsubsection{Comparison with \F}\label{sec:comparison-f}

In the dataset from prior works~\cite{tsbport,fixmorph},
\sys achieves notable performance enhancements
across all backporting types when compared to \F.
For instance,
%
in Type-\RomanNumeralCaps{4} patches,
\F demonstrates a modest success rate of 8.70\%
while \sys improves it to 54.49\%,
showcasing its resilience in addressing complex backporting scenarios.
The performance advantage of \sys is
more remarkable in our dataset for C test cases.
Across all types,
\F achieves a success rate of 4.34\%,
\sys demonstrates superior efficacy with a success rate of 60.87\%.

We analyzed the failure cases of \F within our dataset.
Firstly,
\F enforces stricter constraints by
supporting only C source files while excluding C header files.
This limitation is explicitly stated in the paper and
is further confirmed by its implementation,
which resulted in 14 failed cases.
Among other failed cases,
\F failed to produce results in 33 instances,
primarily because it raised errors
that caused the transformation process to terminate prematurely.
In seven of these failures,
\F was unable to identify the correct patch locations in the target version.
In the remaining cases,
its AST-based approach was unable to handle complex changes,
such as code syntax restructuring and
module-level semantic modification,
resulting in failures during the AST transformation process.

\F generated a total of 22 patches out of which 19 are incorrect,
and they exhibit several issues.
These patches contained syntax errors,
such as disordered statements or
failure to utilize functions or variables compatible with target versions.
Furthermore,
\F struggled to accurately migrate patches involving changes in semantics.
Moreover, the complexity of the code plays a crucial role in the performance of \F.
The average number of modified lines for successfully backported patches is 1.3,
compared to 40.1 for failed patches.
This difference indicates that
\F struggles with cases involving
intricate logic and extensive code modifications.

\subsubsection{Comparison with \T}\label{sec:comparison-t}

In the dataset from prior works \cite{tsbport,fixmorph},
\sys achieves comparable results to \T on simpler patch types but
excels in handling more complex types,
with success rates of 80.43\% and 54.49\% on
Type-\RomanNumeralCaps{3} and Type-\RomanNumeralCaps{4}, respectively,
compared to 75.00\% and 46.38\% in \T.
On our dataset,
the advantage of \sys becomes even more evident.
It achieves 100\% accuracy on Type-\RomanNumeralCaps{2} cases and
significantly outperforms \T on
Type-\RomanNumeralCaps{3} (72.41\% vs. 48.28\%) and
Type-\RomanNumeralCaps{4} (44.12\% vs. 14.71\%). 
%

In addition,
\T requires manual designation of the target file for each hunk and
we provide it with the ground truth in the experiments,
which favors \T.
In particular,
\T does not handle the issue of mismatched target and original patch files,
which is the main challenge in Type-\RomanNumeralCaps{2}
(as these patches requires no transformation at all).
To address this evaluation bias,
we limit files from the original patch as the target in $\T^*$.
\sys outperforms $\T^*$ by 17\% overall.

The underperformance of \T on Type-\RomanNumeralCaps{3} and \RomanNumeralCaps{4} cases in our dataset stems from several issues.
For Type-\RomanNumeralCaps{3},
which involves 15 failure cases,
the primary reason is erroneous symbol importing.
\T often struggles to resolve missing symbols,
such as function or global variable declarations in header files,
during patch migration.
This limitation frequently leads to failed patch applications or
introduction of additional header files imports,
accounting for errors in 11 out of the 15 cases.
These extraneous header file imports are typically
migrated from the header files in the original version,
but these headers do not exist in the target version,
causing compilation error.

Type-\RomanNumeralCaps{4} encompasses 29 failure cases,
which can be attributed to two primary issues.
The first issue is \T's inadequate handling of module-level modifications,
such as updates to struct fields or the introduction of new functions.
This deficiency leads to incorrect mappings,
accounting for 7 out of the 29 failures.
The second issue is the complexity of the code itself.
This is evidenced by a notable difference
in the average number of modified lines:
13.8 for successful patches versus 35.7 for failed ones.
The contrast highlights \T's difficulty
in dealing with complex logic and extensive modifications.
%
%
Note that \sys faces the same difficulty as well---%
more complicated patches are harder to backport---%
but with a higher bar
since the average number of modified lines
for successful patches reached 20.1,
while for failed patches, it averaged 51.0.

\subsubsection{Generalizability over other programming languages}\label{sec:generalizability}
\sys demonstrates strong programming language generalizability. Unlike \F and \T, which are coupled with C only, \sys consistently performs well across multiple programming languages. Notably, adapting \sys from C to C++ and Go required no modification, highlighting its flexibility. These results underscore \sys's potential for robust generalizability and its capability to be effectively extended to support a broader range of programming languages. \T and \F required predefined transformation rules based on expert knowledge, whether for syntax or semantic matching, and were tightly coupled with specific programming languages. In contrast, \sys relies on a series of text-based tools (e.g., GitUtils) for localization and leverages the code generation capabilities of LLM to perform patch transformation. This allows \sys to be language-agnostic.

However, \sys's performance on Type-\RomanNumeralCaps{4} C++ cases is significantly lower, with a success rate of only 5 out of 17 cases (29\%) compared to at least 40\% in other languages. This discrepancy may partly stem from the small dataset, which could introduce statistical bias. Another contributing factor is the complexity of required modifications: 11 out of 12 C++ failure
cases necessitate around 100 lines of changes, whereas most C cases require fewer than 50 lines.

\begin{table}\small
    \centering
    \begin{tabular}{lcccc}
    \toprule
    \textbf{Model} &
    \textbf{Type-\RomanNumeralCaps{2}} &
    \textbf{Type-\RomanNumeralCaps{3}} &
    \textbf{Type-\RomanNumeralCaps{4}} &
    \textbf{Total} \\
    \midrule
    GPT-5  & 19/19 & 42/49 & 31/78 & 92/146 \\
    GPT-4o & 19/19 & 38/49 & 34/78 & 91/146 \\
    Gemini-2.5-Flash & 19/19 & 37/49 & 27/78 & 82/146 \\
    DeepSeek-v3 & 18/19 & 33/49 & 25/78 & 76/146 \\
    Llama-3.3 & 17/19 & 7/49 & 0/78 & 24/146\\
    \bottomrule
    \end{tabular}
    \caption{\textbf{\sys Performance based on Different Large Language Model.}}
    \label{table:models}
    \vspace{-2ex}
\end{table}

\begin{table*}[t]\small
\centering
\tabcolsep=3.5pt
\begin{tabular}{l|ccccccc}
\toprule
\textbf{Line Range} & [1, 5] & [6, 10] & [11, 30] & [31, 50] & [51, 100] & [101, 200] & [200, $\infty$) \\
\midrule
\noalign{\vspace{-0.6ex}}
\multicolumn{8}{c}{\scriptsize\textit{Prior works \cite{fixmorph,tsbport}}} \\[-0.7ex]
\midrule
\sys        & 744/787 (94.5\%) & 339/373 (90.9\%) & 349/401 (87.0\%) & 100/131 (76.3\%) & 68/92 (73.9\%) & 16/24 (66.7\%) & 2/7 (28.6\%) \\
$\T^*$      & 537/787 (68.2\%) & 274/373 (73.5\%) & 297/401 (74.0\%) & 101/131 (77.1\%) & 59/92 (64.1\%) & 20/24 (83.3\%) & 2/7 (28.6\%) \\
\F          & 337/787 (42.8\%) & 85/373 (22.8\%) & 23/401 (5.7\%) & 2/131 (1.5\%) & 0/92 (0\%) & 0/24 (0\%) & 0/7 (0\%) \\
\midrule
\noalign{\vspace{-0.6ex}}
\multicolumn{8}{c}{\scriptsize\textit{Our Dataset (C)}} \\[-0.7ex]
\midrule
\sys        & 10/12 (83.3\%) & 8/10 (80.0\%) & 14/21 (66.7\%) & 5/11 (45.5\%) & 3/7 (42.8\%) & 1/3 (33.3\%) & 2/5 (40.0\%) \\
$\T^*$      & 6/12 (50.0\%) & 4/10 (40.0\%) & 4/21 (19.0\%) & 3/11 (27.3\%) & 4/7 (57.1\%) & 0/3 (0\%) & 0/5 (0\%) \\
\F          & 3/12 (25.0\%) & 0/10 (0\%) & 0/21 (0\%) & 0/11 (0\%) & 0/7 (0\%) & 0/3 (0\%) & 0/5 (0\%) \\
\bottomrule
\end{tabular}
\caption{\textbf{Performance Analysis by Patch Size.} Success rates of different tools across varying patch sizes (measured in modified lines) on prior datasets from and our new C dataset. Numbers indicate successful patches out of total attempts with success percentages in parentheses.}
\label{tab:patchsize}
\end{table*}

\subsubsection{Effectiveness of other large language models}

As shown in \autoref{table:models}, 
\sys demonstrates varying performance across different language models. GPT-5 achieves the best overall performance with 92/146 (63.0\%) success rate, followed closely by GPT-4o with 91/146 (62.3\%). Gemini-2.5-Flash delivers solid performance with 82/146 (56.2\%) overall success, while DeepSeek-v3 shows moderate effectiveness at 77/146 (52.7\%).
This decline in Llama 3.3 performance is primarily attributed to Llama 3.3's limited function-calling capabilities, which are critical for retrieving information in the backporting task. Notably, Llama 3.3 often aborts processes prematurely when faced with insufficient information, instead of attempting additional tool invocations. Moreover, while GPT-5 consistently executes tool calls accurately, Llama 3.3 frequently generates function-calling messages with incorrect formats or parameters, severely compromising its effectiveness. The Berkeley Function Calling Leaderboard \cite{berkeley-function-calling-leaderboard} further validates these other models' superior performance compared to Llama 3.3 in function-calling tasks.
While DeepSeek-v3 demonstrates better performance in tool invocation (compared to Llama 3.3), its accuracy still falls short of the other models due to its limitations in code comprehension and contextual understanding.


\subsubsection{Impact of Patch Size Distribution}

To better understand the relationship between patch size and tool effectiveness, we analyze the distribution of patch lengths in both prior datasets and our constructed dataset. The results consistently demonstrate that the effectiveness of all tools declines as patch length increases. This trend highlights two key challenges: (1) larger hunks are inherently more difficult to handle, and (2) longer patches increase the likelihood of encountering at least one challenging hunk, thus lowering the overall success rate. \autoref{tab:patchsize} summarizes the performance of \sys, $\T^*$, and \F across different patch length ranges, combining both prior datasets and our dataset. Overall, all tools exhibit a monotonic decline in success rate as patch length increases. For instance, on prior datasets, \sys achieves a success rate of 94.5\% on patches of 1--5 lines, but the rate drops to 66.7\% for patches of 100--200 lines. \F is particularly sensitive to patch length, with effectiveness approaching zero beyond 30 lines, while {$\T^*$} shows a less stable trend and also struggles on very long patches. 

\subsection{Ablation Study (RQ2)}

We conducted an ablation study, selectively disabling each of \sys's well-designed tools to measure their impact on patch backporting performance. With each case requiring four evaluations under this protocol, testing a dataset of approximately 2000 cases leads to prohibitive monetary costs, estimated in the thousands of dollars. Therefore, to quantify each tool's contribution to the LLM's backporting capabilities under these constraints, our analysis only utilized our dataset.


\begin{table}[t]\small
    \centering
    \begin{tabular}{l|cccc}
    \toprule
    \textbf{Configuration} & \textbf{Type-II} & \textbf{Type-III} & \textbf{Type-IV} & \textbf{Total} \\
    \midrule
    w/o GitUtils    & 94.74\% & 75.51\% & 32.05\% & 54.79\% \\
    w/o Locate      & 94.74\% & 69.39\% & 38.46\% & 56.16\% \\
    w/o Loc \& Git  & 84.21\% & 59.18\% & 25.64\% & 44.52\% \\
    w/o AutoFix     & 100\%   & 65.31\% & 30.77\% & 51.37\% \\
    w/o Compile     & 100\%   & 69.39\% & 25.64\% & 50.00\% \\
    \midrule
    \textbf{\sys}   & \textbf{100\%} & \textbf{77.55\%} & \textbf{43.59\%} & \textbf{62.33\%} \\
    \bottomrule
    \end{tabular}
    \caption{\textbf{Ablation Study of \sys.} w/o = without. GitUtils, Locate (LocateSymbol) and Compile (CompileTest) are three tools in \sys. AutoFix refers to the patch correction mechanism.}
    \label{tab:ablation}
\end{table}

The results, shown in \autoref{tab:ablation}, highlight the contributions of each tool. Firstly, we assess \textit{GitUtils} and \textit{LocateSymbol} tools that contribute to patch localization. Though these two tools cooperate to achieve better performance, they could also work independently. Disabling \textit{GitUtils} caused a sharp drop in overall performance, particularly in Type-\RomanNumeralCaps{4} cases, where the accuracy plummeted to 32.05\%. This outcome underscores the importance of \textit{GitUtils} for handling complex relationships between commits. Similarly, removing the \textit{LocateSymbol} tool led to reduced performance, especially in Type-\RomanNumeralCaps{3} cases, where accuracy fell to 69.39\%, demonstrating its critical role in identifying symbol locations. Removing both tools simultaneously results in the most severe degradation, with overall accuracy dropping to 44.52\%, confirming their synergistic effect in patch localization.

Secondly, we assess AutoFix mechanism in \textit{ApplyHunk} and \textit{CompileTest} tools that contribute to patch transformation. Disabling AutoFix caused performance to degrade across all types, with Type-\RomanNumeralCaps{3} and Type-\RomanNumeralCaps{4} cases being particularly affected, showing accuracies of 65.31\% and 30.77\%, respectively. This result underscores AutoFix’s critical role in refining intermediate patch quality. Removing \textit{CompileTest} also led to performance drops across all case types, reflecting its importance in final patch combination and integration. Overall, the complete system, \sys, achieves the best performance, excelling in challenging scenarios such as Type-\RomanNumeralCaps{4} cases with 43.59\% accuracy and achieving a total accuracy of 62.33\%. These results confirm that the integration of all components is essential for optimal system performance.

\begin{table}[t]\small
    \centering
    \tabcolsep=7.3pt
    \begin{tabular}{l|cccc}
        \toprule
            \multirow{2}{*}{\textbf{Type}} & \multicolumn{2}{c}{\textbf{Avg. Token}} & \multirow{2}{*}{\textbf{\begin{tabular}[c]{@{}c@{}}Avg.\\ 
            \$ Cost\end{tabular}}}  &
            \multirow{2}{*}{\textbf{\begin{tabular}[c]{@{}c@{}}Avg.\\ 
                Time (s) \end{tabular}}}   \\
            & \textbf{\# Input} & \textbf{\# Output} & \\
        \midrule
            Type-\RomanNumeralCaps{1} & 20,152 & 1,057 & 0.06 &  53.22 \\
            Type-\RomanNumeralCaps{2} & 31,831 & 2,180 & 0.10 & 100.25 \\
            Type-\RomanNumeralCaps{3} & 57,146 & 4,169 & 0.19 & 193.66 \\
            Type-\RomanNumeralCaps{4} & 89,804 & 4,776 & 0.27 & 222.52 \\
        \midrule
            \textbf{Total} & 60,649 & 3,589 & 0.19 & 166.58  \\
        \bottomrule
    \end{tabular}
    \caption{\textbf{Average Time \& Money Cost of \sys.}}
    \label{tab:cost}
\end{table}

\subsection{Efficiency Evaluation (RQ3)}

The efficiency of \sys~is evaluated on the whole dataset in terms of average token usage, monetary cost, and processing time, as shown in \autoref{tab:cost}. Across different input-output scenarios (Type-\RomanNumeralCaps{1} to Type-\RomanNumeralCaps{4}), the average time and cost scale proportionally with the complexity of the task, with larger input and output sizes leading to higher resource consumption. As the complexity of patches increases (from Type-\RomanNumeralCaps{1} to Type-\RomanNumeralCaps{4}), the LLM requires more information queries and performs more reasoning, leading to higher costs. The total average input token count is 60,649, producing an average of 3,589 output tokens, with a cost of \$0.19 and a processing time of 166.58 seconds. This balance between cost-effectiveness and processing speed demonstrates the practical scalability of \sys~for various levels of task complexity, maintaining reasonable efficiency even for the most demanding scenarios.

\begin{table}[t]
    \centering
    \tabcolsep=3.4pt
    \begin{tabular}{llcccc}
    \toprule
    \textbf{CVE ID} & \textbf{Version} & \textbf{Fix Date} & \textbf{Type} & \textbf{P.G.} \\ 
    \midrule
    CVE-2024-46743 \cite{linux-patch-CVE-2024-46743} & Bionic & 2024/10/15  &  \RomanNumeralCaps{2} & \ding{51} \\
    CVE-2023-51779 \cite{linux-patch-CVE-2023-51779} & Focal & 2024/01/05   &  \RomanNumeralCaps{3} & \ding{51} \\
    CVE-2024-0565  \cite{linux-patch-CVE-2024-0565} & Focal & 2024/01/29    &  \RomanNumeralCaps{3} & \ding{51} \\
    CVE-2024-26922 \cite{linux-patch-CVE-2024-26922} & Focal & 2024/05/21   &  \RomanNumeralCaps{3} & \ding{51} \\
    CVE-2024-43863 \cite{linux-patch-CVE-2024-43863} & Focal & 2024/12/03   &  \RomanNumeralCaps{3} & \ding{51} \\
    CVE-2024-43863 \cite{linux-patch-CVE-2024-43863} & Bionic & N/A         &  \RomanNumeralCaps{3} & \ding{51} \\ 
    CVE-2023-24023 \cite{linux-patch-CVE-2023-24023} & Focal & 2024/03/14   &  \RomanNumeralCaps{4} & \ding{55} \\
    CVE-2023-52752 \cite{linux-patch-CVE-2023-52752} & Bionic & 2024/06/26  &  \RomanNumeralCaps{4} & \ding{55} \\
    CVE-2023-52752 \cite{linux-patch-CVE-2023-52752} & Focal & 2024/06/26   &  \RomanNumeralCaps{4} & \ding{51} \\
    CVE-2023-52752 \cite{linux-patch-CVE-2023-52752} & Jammy & 2024/06/26   &  \RomanNumeralCaps{4} & \ding{55} \\
    CVE-2024-24860 \cite{linux-patch-CVE-2024-24860} & Bionic & 2024/07/09  &  \RomanNumeralCaps{4} & \ding{55} \\
    CVE-2024-24860 \cite{linux-patch-CVE-2024-24860} & Focal & 2024/07/09   &  \RomanNumeralCaps{4} & \ding{55} \\
    CVE-2024-25742 \cite{linux-patch-CVE-2024-25742} & Jammy & 2024/07/02   &  \RomanNumeralCaps{4} & \ding{55} \\
    CVE-2024-26922 \cite{linux-patch-CVE-2024-26922} & Bionic & 2024/05/21  &  \RomanNumeralCaps{4} & \ding{51} \\
    CVE-2024-41066 \cite{linux-patch-CVE-2024-41066} & Focal & 2024/11/25   &  \RomanNumeralCaps{4} & \ding{51} \\
    CVE-2024-41066 \cite{linux-patch-CVE-2024-41066} & Jammy & 2024/11/25   &  \RomanNumeralCaps{4} & \ding{51} \\
    \bottomrule
    \end{tabular}
    \caption{\textbf{Ubuntu Backporting Tasks Results.} \textbf{Version} refers to the target version for Ubuntu backporting tasks. \textbf{Fix Date} represents the actual date of the backport in the real world.}
    \label{tab:unseen}
\end{table}

\subsection{Real-World Applicability Study (RQ4)}

To evaluate the practical applicability 
of \sys, we conducted studies across two prevalent and challenging real-world backporting scenarios: (1) porting patches from the mainline Linux kernel to a LTS version, and (2) propagating patches from an LTS version to downstream distributions. All patches selected for these scenarios were introduced after the knowledge cutoff of the underlying LLM (GPT-4o), allowing for a robust assessment of \sys's generalization to unseen tasks.

\paragraph{Mainline To LTS} For this scenario, our selection process focused on challenging CVE patches (2024-2025, sourced from Linux vulns~\cite{vuls.git})—specifically those that had previously failed backporting from mainline to Linux 6.1-stable and lacked manual conflict resolution. This rigorous filtering yielded a focused set of just 18 such cases. \sys successfully handled 9 of these.
%
Critically, all 9 generated patches were subsequently reviewed and merged into the 6.1-stable branch by kernel maintainers\cite{linux-merged-patch-CVE-2025-21816,linux-merged-patch-CVE-2024-26618,linux-merged-patch-CVE-2024-26807,linux-merged-patch-CVE-2024-36903,linux-merged-patch-CVE-2024-36927,linux-merged-patch-CVE-2024-53209,linux-merged-patch-CVE-2024-56758,linux-merged-patch-CVE-2024-57945,linux-merged-patch-CVE-2025-21645}. For the remaining 9 cases, 7 failed due to significant structural code changes, and 2, though directly cherry-pickable, exhibited regressions post-application~\cite{CVE-2024-26889-Patch-Revert}. In contrast, \T handled only 2/18 cases (11.1\%), underscoring \sys's improved applicability for complex mainline-to-LTS backports.

\paragraph{LTS To Downstream} We evaluated \sys on 16 patch pairs (10 CVEs) for Ubuntu (targeting Jammy, Focal, Bionic versions)~\cite{ubuntu}. All patches were introduced after October 2023 to test generalization when porting from LTS sources to downstream targets. \sys successfully backported 10 out of 16 tasks (as shown in \autoref{tab:unseen}). This demonstrates \sys's robust generalization for LTS-to-downstream backporting of unseen patches.

\subsection{Case Studies}

In this part, we present two cases to illustrate how the design of \sys effectively supports LLM reasoning.

\paragraph{CVE-2020-22030} In both versions, the patch introduces a length check before usage as the core fix. However, as code evolves across versions, the function and struct names, as well as their usages in the patch, have changed. As a result, beyond resolving contextual conflicts, the added logic in the patch also requires careful adaptation to align with the target version’s codebase, which poses a significant challenge for automated backporting tools. Specifically, the function \texttt{ff\_inlink\_queued\_samples} was replaced by \texttt{ff\_framequeue\_queued\_samples}, and the field \texttt{inputs} was used differently. In this case, the \textit{History} component in \sys's \textit{GitUtils} module effectively provided relevant historical commit information to the LLM, enabling it to reason about and adapt the necessary changes. In contrast, due to the lack of historical context, \T reused the mainline implementation without adapting to version-specific changes, leading to an incorrect backport.

\paragraph{CVE-2023-41175} The added lines in the patch also require adaptation to fit the target version. Specifically, the macro \texttt{UINT\_MAX} does not exist in the target version. However, this issue cannot be inferred from the patch’s code context or its historical modifications, and can only be revealed through compilation and testing. \T, which performs no post-generation verification (including compilation), produces a ``seemingly correct'' patch that ultimately fails. In contrast, \sys adopts a two-stage design that effectively eliminates hard-to-infer transformations during backporting. Its iterative process, compilation testing and patch correction based on observed errors, also closely aligns with real-world backporting practices.
A detailed analysis of the failed cases can be found in~\ref{sec:case_study}.

%% file: sections/7_discussion.tex
\section{Discussion and Limitations}\label{sec:discussion}

In this section, we discuss the limitations of \sys.

\paragraph{Results Interpretation} While \sys demonstrates promising results, it currently cannot consistently provide reliable interpretations of backported patches to explain their correctness. This limitation arises because LLMs are statistical models of syntactic representation. Although they can generate syntactically correct code, they lack the ability to provide a robust semantic understanding. Therefore, we recommend that \sys users treat backported patches and outputs as suggestions, carefully reviewing the patches manually before applying them to the target codebase.

\paragraph{Context Length} 
One of the key limitations of LLMs is their restricted context length, which can hinder performance when the input exceeds a certain size. Research \cite{levy2024same} has shown that the effectiveness of LLMs decreases as the length of the input increases, due to their inability to process long-range dependencies effectively. In \sys, this issue is mitigated by processing the hunks of the original patch separately during the first stage, allowing each part to be handled within the context length constraints. However, this strategy comes with its own challenges, as it can result in a loss of crucial context from the original patch, potentially affecting the accuracy of the backporting process.

%% file: sections/8_related.tex
\section{Related Work}

\paragraph{Program Repair} Automated program repair (APR) aims to automatically fix bugs in software systems, sharing a goal similar to backporting but operating without access to the original patch. Among the various tasks in APR, patch generation has received the most attention. This task involves taking a buggy code snippet as input and producing a patch to fix the bug. Many approaches leveraging deep learning \cite{fu2022vulrepair,chen2022neural,guo2020graphcodebert} and large language models (LLMs) \cite{patchagent,wei2023copiloting,xia2022less,xia2023keep} have shown promising results in this area. A prerequisite task for APR is fault localization (FL), which identifies the buggy code snippet. This process resembles \sys, where LLMs are used to determine the corresponding patch location of a hunk. However, FL in APR is often more challenging due to the absence of the original patch. Current FL methods are predominantly statistical \cite{shen2021localizing,blazytko2020aurora,park2023benzene,xu2024racing}, scoring elements in the program to analyze root causes. Another critical step in APR is patch validation, which is typically conducted through either manual review \cite{wu2023mitigating} or automated testing \cite{pearce2023examining}. In this work, we primarily rely on manual efforts to validate the backported patches.

\paragraph{LLM for SE} LLM have become powerful tools for various software engineering (SE) tasks, such as vulnerability detection~\cite{sun2024llm4vuln,ding2024vulnerability}, code translation~\cite{pan2023lost, lachaux2020unsupervised}, program repair~\cite{xia2022less,first2023baldur,pearce2023examining,ahmad2023fixing}, and unit test generation~\cite{alshahwan2024automated,alshahwan2024assured}. To assess their effectiveness in real-world applications, benchmarks like SWE-bench~\cite{jimenez2023swe} have been developed. SWE-bench consists of many GitHub issues and their corresponding pull requests from widely-used Python repositories, providing a challenging dataset for evaluating how well models can generate patches to resolve specific codebase issues. Several notable tools have been evaluated using SWE-bench. SWE-Agent~\cite{yang2024swe} leverages LLMs to autonomously tackle GitHub issues, utilizing a custom Agent-Computer Interface (ACI) to improve the model's ability to navigate, edit, and test code within repositories. Similarly, AutoCodeRover~\cite{zhang2024autocoderover} integrates LLMs with advanced code search mechanisms, leveraging program structures such as abstract syntax trees to enhance bug fixing and feature implementation. Building on strong capabilities of LLMs, our work uses these models to facilitate patch backporting between different versions.

%% file: sections/9_conclusion.tex
\section{Conclusion}

Patch backporting is a complex and labor-intensive task in the real world, which increases the burden of project maintainers. In this work, we introduced \sys, an LLM-based tool designed for end-to-end automated patch backporting. Using the modification history effectively in Git, \sys employs a combined approach to achieve precise localization, guiding the LLM to make inferences based on the provided information, thus simulating the expert backporting process. Our extensive evaluation on diverse datasets demonstrate that \sys significantly outperforms existing automated patch backporting tools. Evaluations on multi-language, multi-project, and more complex patch types also demonstrate \sys's ability to handle diverse scenarios and its versatility.

%





%% file: sections/10_appendix.tex
\section{}

\begin{table}[!t]\small
    \centering
    \tabcolsep=4pt
    \begin{tabular}{c|c|c|c|c|c}
    \toprule
    \textbf{Language} & \textbf{Project} & 
    \textbf{Type-\RomanNumeralCaps{2}} & 
    \textbf{Type-\RomanNumeralCaps{3}} & 
    \textbf{Type-\RomanNumeralCaps{4}} & 
    \textbf{Total} \\ \midrule
    \multirow{14}{*}{C} 
    & Linux                         &   6   &   4   &   11   &   21                    \\
    & FFmpeg                        &   0   &   5   &   7   &   12                    \\
    & redis                         &   0   &   6   &   4   &   10                    \\
    & libtiff                       &   0   &   3   &   3   &   6                    \\
    & cpython                       &   0   &   2   &   2   &   4                    \\
    & glibc                         &   0   &   3   &   1   &   4                    \\
    & gnupg                         &   0   &   1   &   2   &   3                    \\
    & krb5                          &   0   &   1   &   1   &   2                    \\
    & libwebp                       &   0   &   1   &   1   &   2                    \\
    & FreeRDP                       &   0   &   1   &   0   &   1                    \\
    & OpenSSL                       &   0   &   0   &   1   &   1                    \\
    & libpng                        &   0   &   1   &   0   &   1                    \\
    & php                           &   0   &   1   &   0   &   1                    \\
    & sqlite                        &   0   &   0   &   1   &   1                    \\
    \midrule
    \multirow{7}{*}{C++} 
    & MongoDB                       &   0   &   3   &   7   &   10                    \\
    & Chromium                      &   0   &   2   &   5   &   7                    \\
    & envoy                         &   0   &   4   &   1   &   5                    \\
    & electron                      &   0   &   1   &   1   &   2                    \\
    & qt                            &   0   &   1   &   1   &   2                    \\
    & grpc                          &   0   &   0   &   1   &   1                    \\
    & v8                            &   0   &   0   &   1   &   1                    \\
    \midrule
    \multirow{13}{*}{Go} 
    & golang                        &   6   &   2   &   7   &   15                    \\
    & Argo CD                       &   3   &   1   &   7   &   11                    \\
    & consul                        &   3   &   1   &   2   &   6                    \\
    & containerd                    &   0   &   0   &   4   &   4                    \\
    & nomad                         &   1   &   1   &   1   &   3                    \\
    & mattermost                    &   0   &   0   &   2   &   2                    \\
    & moby                          &   0   &   1   &   1   &   2                    \\
    & Valut                         &   0   &   1   &   0   &   1                    \\
    & cilium                        &   0   &   0   &   1   &   1                    \\
    & darp                          &   0   &   0   &   1   &   1                    \\
    & etcd                          &   0   &   1   &   0   &   1                    \\
    & go-jose                       &   0   &   1   &   0   &   1                    \\
    & Grafana                       &   0   &   0   &   1   &   1                    \\
    \midrule
    \textbf{Total} & \multicolumn{1}{c|}{/} & 19 & 49 & 78 & 146\\
    \bottomrule
    \end{tabular}
    
    \caption{\textbf{Overview of Our Dataset.}}
    \label{tab:dataset}
\end{table}

\subsection{Dataset}\label{appdendix:dataset}

\autoref{tab:dataset} presents an overview of the dataset collected by us. It includes the number of patches for each type and the projects from which they were collected. The dataset includes a total of 34 projects. During the patch collection process, we also gathered the compilation scripts required for each patch, to be used in subsequent testing.

\begin{figure}[t]
\vspace{-1.5ex}
\begin{lstlisting}[language=diff]
@@ -4397,8 +4398,15 @@ __cgroup_procs_write(
  spin_unlock_irq(&css_set_lock);
+ saved_cred = override_creds(of->file->f_cred);
  ret = cgroup_attach_permissions(src_cgrp, 
  dst_cgrp, of->file->f_path.dentry->d_sb, ...);
+ revert_creds(saved_cred);
  if (ret)
      goto out_finish;

@@ -4440,9 +4449,15 @@ cgroup_threads_write(
  spin_unlock_irq(&css_set_lock);
+ saved_cred = override_creds(of->file->f_cred);
  ret = cgroup_procs_write_permission(src_cgrp,
      dst_cgrp, of->file->f_path.dentry->d_sb);
+ revert_creds(saved_cred);
  if (ret)
      goto out_finish;
\end{lstlisting}
\vspace{-0.5ex}
\captionof{listing}{\textbf{CVE-2021-4197.} Both the original \cite{linux-mainline-patch-CVE-2021-4197} and target \cite{linux-stable-patch-CVE-2021-4197} version patches  include the first hunk, but only the target version requires modification of the second hunk.}
\label{lst:CVE-2021-4197}
\end{figure}

\begin{figure}[h]
\centering
\includegraphics[width=0.85\columnwidth]{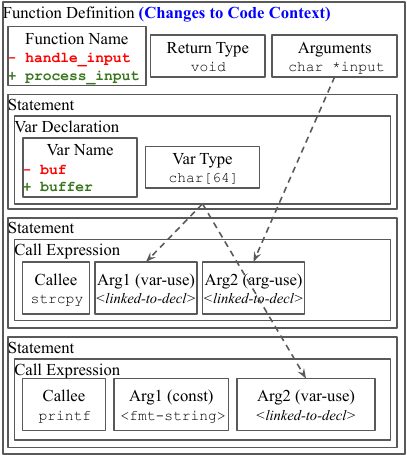}
\caption{The AST representation of
how $L_o$ changes to $L_n$ around $\Delta P$
in \lstref{lst:bg-patch-pn}: two symbol renamings.}
\label{fig:bg-ast-po-pn}
\vspace{-2ex}
\end{figure}

\begin{figure}[h]
\centering
\includegraphics[width=0.85\columnwidth]{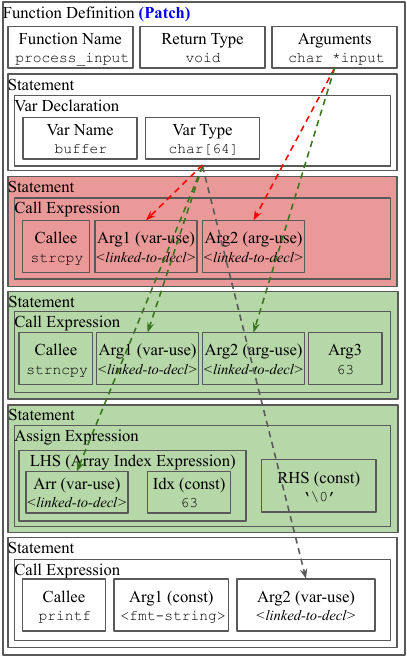}
\caption{The AST representation of
$\Delta P$ shown in Listing~\ref{lst:bg-patch-pn}.
Red marks deletion while green marks addition.}
\label{fig:bg-patch-pn-ast}
\vspace{-2ex}
\end{figure}

\subsection{Failure Case Studies}\label{sec:case_study}


In this section, We describe the main reasons behind the failures of \sys and present several case studies to illustrate the underlying causes of these failures.

From a high-level perspective, the reasons for failure can be summarized into three points: 1) The target version contains additional vulnerable code compared to the original version, which requires fixing; 2) The hunk-by-hunk backport approach overlooks potential dependencies between hunks. 3) \sys only backports the current patch and does not incorporate its prerequisite commits. The following cases describe these three points in detail.


\paragraph{CVE-2021-4197 \cite{linux-mainline-patch-CVE-2021-4197,linux-stable-patch-CVE-2021-4197}} CVE-2021-4197 highlights a common failure reason in \sys's backporting process, as demonstrated in \lstref{lst:CVE-2021-4197}. The patch in the original version \cite{linux-mainline-patch-CVE-2021-4197} includes the first hunk, which introduces proper credential handling through \texttt{override\_creds} and \texttt{revert\_creds} for \texttt{cgroup\_attach\_permissions}. While \sys successfully applies this modification to the corresponding code in the target version, it fails to address a similar vulnerability in the second hunk involving \texttt{cgroup\_procs\_write\_permission}. Both code segments require identical security enhancements to handle credentials correctly \cite{linux-stable-patch-CVE-2021-4197}. The oversight leaves the second vulnerability unpatched and cause \sys generates an incomplete patch. In the mainline version, the second segment does not exist since \texttt{cgroup\_threads\_write} is implemented with \texttt{\_\_cgroup\_procs\_write}.





\paragraph{CVE-2019-16232 \cite{linux-mainline-patch-CVE-2019-16232}} 
CVE-2019-16232 illustrates a failure in \sys's backporting process. In the original version, the patch consists of two related hunks: the first introduces the error handling statement \texttt{goto err\_queue;}, and the second defines the behavior of the \texttt{err\_queue} label. During backporting, \sys replaces \texttt{err\_queue} with \texttt{out} in the first hunk because it cannot find the \texttt{err\_queue} label in the target version’s context. When processing the second hunk, \sys directly adds the \texttt{err\_queue} label as in the original patch, but the label does not integrate correctly because it was not previously established. We believe this failure is due to \sys's hunk-by-hunk approach, which fails to account for dependencies between hunks.

\paragraph{CVE-2022-41720\cite{golang-newer-patch-CVE-2022-41720, golang-older-patch-CVE-2022-41720}} CVE-2022-41720 is a vulnerability in the Golang project that highlights a limitation in \sys’s backporting process. The patch in the master branch \cite{golang-newer-patch-CVE-2022-41720} revises the \texttt{join} function to address the issue; however, this function does not exist in the target version. To address this, the patch for the target version \cite{golang-older-patch-CVE-2022-41720} first introduces the \texttt{join} function before modifying it. \sys fails to backport the patch because it cannot locate the \texttt{join} function in the target version and incorrectly assumes that no backporting is needed. This failure stems from \sys's inability to follow a common backporting practice where developers include prerequisite commits to adjust the context before applying the actual patch. Addressing this limitation in future work will enable \sys to better handle complex backporting scenarios, aligning more closely with experienced developer practices.

\subsection{Illustration of AST-based backporting}

\autoref{fig:bg-ast-po-pn} illustrates \F's context matching process using AST. Even though this example involves the renaming of functions and variables, their underlying syntactic structure remains identical at the AST level. This allows \F to effectively handle such superficial textual variations within the code context.

\autoref{fig:bg-patch-pn-ast}, on the other hand, depicts \F's transformation stage. In this visualization, red highlights AST structures identified as identical and thus marked for deletion, while green denotes newly added AST structures. Finally, nodes corresponding to variables that have undergone changes are updated with their new names (e.g., \texttt{buf}).

\begin{figure}[!t]
    \centering
\begin{tcolorbox}[colback=gray!5!white,colframe=gray!75!black, title=Stage-1 Prompt]
\small
Your \textbf{TASK} is to backport a patch fixing a vuln from a original version of the software to an target version step by step. The project is \{project\_url\}. For the ref \{original\_patch\_parent\}, the patch below is merged to fix a security issue:

\{original\_patch\}

I want to backport it to ref \{target\_patch\_parent\}. To assist you in reviewing the relevant code, I have provided the following code blocks from the target version that closely match the current patch location:

\{similar\_block\}

Your \textbf{WORKFLOW} should be: ......
\end{tcolorbox}
\begin{tcolorbox}[colback=gray!5!white, colframe=gray!75!black, title=Stage-2 Prompt]
    \small
Your \textbf{TASK} is to validate and revise the patch until it is successfully backported to the target version and really fixes the vulnerability.

Below is the patch you need to backport:

\{new\_patch\}

According to the patch above, I have formed a patch that can be applied to the target version: 

\{complete\_patch\}

Now, I have tried to compiled the patched code, the result is:
\{compile\_ret\}

You can validate the patch with provided tool \textit{validate}. There are some processes to validate if the patch can fix the vulnerability:
......

If the patch can not pass above validation, you need to revise the patch with the help of provided tools. The patch revision \textbf{WORKFLOW} should be: ......
\end{tcolorbox}
    \caption{\textbf{User Prompt of \sys}}
    \label{fig:prompt}
\end{figure}

\subsection{User Prompt of \sys}

\autoref{fig:prompt} details the prompts designed for the LLM specifically for patch backporting tasks. For the two identified stages of this process, we have established distinct workflows for the LLM. These initial prompts direct the LLM to invoke appropriate tools according to the specified workflow and subsequently analyze the resulting outputs.

\subsection{Future Directions} 

To address the limitations outlined above, we propose several promising avenues for future research aimed at enhancing \sys's performance. First, we recommend exploring advanced reasoning techniques \cite{yao2022react, yao2024tree}, as well as leveraging fine-tuning methods \cite{chen2021evaluating}, to improve the ability of LLMs to handle the intricate dependencies in code. These techniques could help the models better understand the semantic relationships between different parts of the code, leading to more accurate and reliable backported patches. Second, we suggest investigating improved methods for validating the correctness of generated patches. In our current work, we rely on ground truth data to assess the accuracy of the backported patches. However, in real-world scenarios, ground truth data is often unavailable, making it difficult to automate the validation process. Although directed fuzzing is also an effective approach for validating patches, current directed fuzzing techniques are limited in generalizability and cannot be widely applied across diverse programs.
We believe that the integration of an automatic validation system could not only streamline this process but also provide valuable feedback that could be used to further refine the LLM's ability to generate high-quality backporting patches.

%% file: sections/11_meta_review.tex
\newpage 

\newpage 

\section{Meta-Review}

The following meta-review was prepared by the program committee for the 2026
IEEE Symposium on Security and Privacy (S\&P) as part of the review process as
detailed in the call for papers.

\subsection{Summary}

The paper tackles the issue of patch backporting in open-source projects by introducing PortGPT, an LLM-driven agent that attempts to mirror developer workflows and is equipped with tools to find symbols and relevant code, adapt each hunk, and verify compilation results before combining a final patch. 

\subsection{Scientific Contributions}
\begin{itemize}
\item Addresses a Long-Known Issue
\item Provides a Valuable Step Forward in an Established Field
\end{itemize}

\subsection{Reasons for Acceptance}
\begin{enumerate}
\item The paper proposes the use of augmented, agentic-based backporting to simplify/automate the process of backporting. The used combination of LLMs with traditional program analysis is novel and effective.
\item The evaluation shows that PortGPT is effective even on more complex patches that have syntactic or structural changes, and that, overall, it achieves better performance than other automated tools.
\end{enumerate}

